\PassOptionsToPackage{unicode}{hyperref}
\PassOptionsToPackage{hyphens}{url}
\PassOptionsToPackage{dvipsnames,svgnames,x11names}{xcolor}
\documentclass[
  11pt,
]{article}
\usepackage{amsmath,amssymb}
\usepackage{iftex}
\ifPDFTeX
  \usepackage[T1]{fontenc}
  \usepackage[utf8]{inputenc}
  \usepackage{textcomp} % provide euro and other symbols
\else % if luatex or xetex
  \usepackage{unicode-math} % this also loads fontspec
  \defaultfontfeatures{Scale=MatchLowercase}
  \defaultfontfeatures[\rmfamily]{Ligatures=TeX,Scale=1}
\fi
\usepackage{lmodern}
\ifPDFTeX\else
\fi
\IfFileExists{upquote.sty}{\usepackage{upquote}}{}
\IfFileExists{microtype.sty}{% use microtype if available
  \usepackage[]{microtype}
  \UseMicrotypeSet[protrusion]{basicmath} % disable protrusion for tt fonts
}{}
\usepackage{xcolor}
\usepackage[margin=1in]{geometry}
\usepackage{graphicx}
\makeatletter
\def\maxwidth{\ifdim\Gin@nat@width>\linewidth\linewidth\else\Gin@nat@width\fi}
\def\maxheight{\ifdim\Gin@nat@height>\textheight\textheight\else\Gin@nat@height\fi}
\makeatother
\setkeys{Gin}{width=\maxwidth,height=\maxheight,keepaspectratio}
\makeatletter
\def\fps@figure{htbp}
\makeatother
\providecommand{\tightlist}{%
  \setlength{\itemsep}{0pt}\setlength{\parskip}{0pt}}
\newlength{\cslhangindent}
\newlength{\csllabelwidth}
\newlength{\cslentryspacingunit} % times entry-spacing
\newenvironment{CSLReferences}[2] % #1 hanging-ident, #2 entry spacing
 {% don't indent paragraphs
  \setlength{\parindent}{0pt}
  \ifodd #1
  \let\oldpar\par
  \def\par{\hangindent=\cslhangindent\oldpar}
  \fi
  \setlength{\parskip}{#2\cslentryspacingunit}
 }%
 {}
\usepackage{calc}

\usepackage{accents}
\usepackage{algorithm}
\usepackage{algpseudocode}
\usepackage{amsmath}
\usepackage{bm}
\usepackage{booktabs}
\usepackage{braket}
\usepackage{caption}
\usepackage{epigraph}
\usepackage{extarrows}
\usepackage{float}
\usepackage[bottom,flushmargin,hang,multiple]{footmisc}
\usepackage{graphicx}
\usepackage{hyperref}
\usepackage[export]{adjustbox}
\usepackage{libertine}
\usepackage[libertine]{newtxmath}
\usepackage[mathcal]{euscript}
\usepackage{mathrsfs}
\usepackage{multirow}
\usepackage{pdflscape}
\usepackage{quoting}
\usepackage{scrextend}
\usepackage{setspace}
\usepackage{siunitx}
\usepackage{soul}
\usepackage{subcaption}
\usepackage{threeparttablex}
\usepackage{titlesec}
\usepackage{xltabular}
\usepackage[utf8]{inputenc}

\titleformat{\section} {\normalfont\fontsize{12}{12}\bfseries\itshape} {}{0em}{\centering \MakeUppercase}[\vspace{0.5ex}]
\titleformat{\subsection} {\normalfont\fontsize{12}{12}\bfseries\itshape} {}{-0.5em}{}[]

\newcolumntype{b}{X}
\newcolumntype{s}{>{\hsize=.5\hsize}X}

\ifLuaTeX
  \usepackage{selnolig}  % disable illegal ligatures
\fi
\IfFileExists{bookmark.sty}{\usepackage{bookmark}}{\usepackage{hyperref}}
\IfFileExists{xurl.sty}{\usepackage{xurl}}{} % add URL line breaks if available
\hypersetup{
  pdftitle={Safeguarding Marketing Research: The Generation, Identification, and Mitigation of AI-Fabricated Disinformation},
  colorlinks=true,
  linkcolor={Maroon},
  filecolor={Maroon},
  citecolor={Blue},
  urlcolor={blue},
  pdfcreator={LaTeX via pandoc}}

\title{Safeguarding Marketing Research: The Generation, Identification,
and Mitigation of AI-Fabricated Disinformation}
\author{}
\date{\vspace{-2.5em}}

\begin{document}
\maketitle

\quotingsetup{font={itshape}, leftmargin=1em, rightmargin=1em, vskip=1ex}

\begin{center}
\author
{Anirban Mukherjee$^{1\ast}$\\
\medskip
\normalsize{$^{1}$Samuel Curtis Johnson Graduate School of Management, Cornell University,}\\
\normalsize{Sage Hall, Ithaca, NY 14850, USA}\\
\smallskip
\normalsize{$^\ast$To whom correspondence should be addressed; E-mail: am253@cornell.edu.}\\
}
\end{center}
\medskip
\singlespacing
\begin{center}
\textbf{Abstract}
\end{center}
\singlespacing

\noindent Generative AI has ushered in the ability to generate content
that closely mimics human contributions, introducing an unprecedented
threat: Deployed en masse, these models can be used to manipulate public
opinion and distort perceptions, resulting in a decline in trust towards
digital platforms.

This study contributes to marketing literature and practice in three
ways. First, it demonstrates the proficiency of AI in fabricating
disinformative user-generated content (UGC) that mimics the form of
authentic content. Second, it quantifies the disruptive impact of such
UGC on marketing research, highlighting the susceptibility of analytics
frameworks to even minimal levels of disinformation. Third, it proposes
and evaluates advanced detection frameworks, revealing that standard
techniques are insufficient for filtering out AI-generated
disinformation.

We advocate for a comprehensive approach to safeguarding marketing
research that integrates advanced algorithmic solutions, enhanced human
oversight, and a reevaluation of regulatory and ethical frameworks. Our
study seeks to serve as a catalyst, providing a foundation for future
research and policy-making aimed at navigating the intricate challenges
at the nexus of technology, ethics, and marketing.

\begin{center}\rule{0.5\linewidth}{0.5pt}\end{center}


\newpage
\doublespacing

\hypertarget{introduction}{%
\section{Introduction}\label{introduction}}

\noindent ``\ldots{} the benefits of generative AI also come with
challenges, and we expect to see attempts from businesses and
individuals to use tools like ChatGPT to manipulate content \ldots{}''

\hfill ---Tripadvisor Transparency Report 2023

User-generated content (UGC), such as reviews and social media posts,
plays a pivotal role in capturing consumer behaviors and preferences
(\protect\hyperlink{ref-chevalier2006effect}{Chevalier and Mayzlin
2006}, \protect\hyperlink{ref-fader2012introduction}{Fader and Winer
2012}, \protect\hyperlink{ref-moe2011value}{Moe and Trusov 2011},
\protect\hyperlink{ref-zhang2020frontiers}{Zhang et al. 2020}). However,
the integrity of insights derived from UGC hinges on the authenticity of
the content. Disinformative\footnote{Misinformation is false or
  misleading information that is spread, regardless of intent.
  Disinformation is the deliberate creation and dissemination of false
  or misleading information with the intent to deceive or manipulate
  others.} practices compromise the value of UGC, eroding trust and
distorting the accuracy of marketing research
(\protect\hyperlink{ref-dellarocas2006strategic}{Dellarocas 2006}).

Alarmingly, fraudulent UGC---encompassing fake reviews and
testimonials---is widespread. Despite over 99.5\% of shoppers consulting
online reviews (\protect\hyperlink{ref-Team_2023}{Team 2023}), up to
42\% of these testimonials may be unreliable
(\protect\hyperlink{ref-News_Lee_2020}{Lee 2020}), misleading millions
of consumers and costing businesses over \$150 billion annually
(\protect\hyperlink{ref-Marciano2021}{Marciano 2021}). Businesses, both
large and small, have been found to fabricate positive endorsements,
despite these practices being illegal and prohibited, damaging the
credibility of major platforms like Google, Amazon, and Yelp
(\protect\hyperlink{ref-Thompson_2023}{Thompson 2023}). Recent
regulations and collaborative initiatives aim to combat this type of
disinformation, including imposing civil penalties of up to \$43,792 per
infraction (\protect\hyperlink{ref-Mark_2021}{Mark S. Goodrich 2021}).
However, the current scale of fraud, coupled with the continuous
evolution of misleading tactics, presents a complex and enduring
challenge (\protect\hyperlink{ref-fradkin2023incentives}{Fradkin and
Holtz 2023}).

The advent of generative artificial research (AI) introduces further
complexity. AI-driven technologies are now extensively used in
applications ranging from copywriting tools such as Jasper, Writer, and
Copy.ai, to visual content platforms like DALL-E 3, Midjourney, Runway
Gen-2, Synthesia, Canva, and Adobe Firefly. This widespread adoption
signals a potential for the mass production of disinformation at
significantly reduced costs, posing new challenges for content
authenticity (\protect\hyperlink{ref-hagendorff2023deception}{Hagendorff
2023}, \protect\hyperlink{ref-park2023ai}{Park et al. 2023}).

A shift towards AI-driven content creation could fundamentally alter the
economics of disinformative content, enabling malicious actors to
manipulate consumer sentiment at scale on major platforms like Amazon,
Yelp, Instagram, and TikTok, and facilitating the widespread
availability of artificially fabricated content
(\protect\hyperlink{ref-weidinger2021ethical}{Weidinger et al. 2021}),
further undermining the reliability of insights from marketing research
(\protect\hyperlink{ref-he2022detecting}{He et al. 2022},
\protect\hyperlink{ref-zellers2019defending}{Zellers et al. 2019}).
Already, anecdotal reports suggest the use of AI by Amazon merchants,
although the specific implications for disinformation are yet to be
fully understood (\protect\hyperlink{ref-Palmer_2023}{Palmer 2023}).

However, even as AI heightens the risk of disinformation, it also offers
the potential for significantly improved detection. AI is at the
forefront of emergent systems for the identification of AI-generated
content, as adopted by major technology companies including Facebook,
LinkedIn, and Walmart (\protect\hyperlink{ref-McCallum_2023}{McCallum
2023}). AI models adeptly recognize linguistic patterns and semantic
text relationships, enabling them to identify contextual
inconsistencies, logical fallacies, and stylistic oddities common in
AI-generated narratives. Thus, advances in AI present challenges but
also opportunities to enhance trust and transparency.

In this paper, we examine the dual nature of AI, as both a tool for
creating and combating disinformation. While existing literature
primarily focuses on the impact of disinformation in the form of fake
reviews on consumer behavior and purchase decisions
(\protect\hyperlink{ref-lappas2016impact}{Lappas et al. 2016},
\protect\hyperlink{ref-luca2016fake}{Luca and Zervas 2016},
\protect\hyperlink{ref-mayzlin2014promotional}{Mayzlin et al. 2014}), we
adopt an alternative perspective: We operate within a framework of
information manipulation and signal jamming, where a predatory firm uses
disinformation to obscure market characteristics, deliberately
misleading rivals into sub-optimal actions.

Such phenomena are termed `astroturfing' in the political science
literature, whereby firms or entities create multiple social media
accounts to simulate grassroots support or opposition to influence
public opinion (\protect\hyperlink{ref-keller2020political}{Keller et
al. 2020}). Astroturfing's marketing roots extend deep, with examples
tracing back to the turn of the 20th century when disposable cup
manufacturers carried out surreptitious public health campaigns to
influence consumer behavior (\protect\hyperlink{ref-lee2010roots}{Lee
2010}) and the formation of the National Smokers Alliance, a public
relations front group funded by the tobacco industry to advocate for
access to tobacco products
(\protect\hyperlink{ref-givel2007consent}{Givel 2007}). In contrast to
these examples, however, AI-facilitated astroturfing is distinguished by
the generation and spread of credible reviews at scale.

Central to our analysis, we investigate whether strategically
disseminated AI-generated marketing disinformation can alter rivals'
marketing research and affect their decision-making. For instance,
Naumzik et al. (\protect\hyperlink{ref-naumzik2022will}{2022}) show that
online reviews can serve as a direct measure of customer satisfaction
and predict business failures. By extension, negative (positive)
disinformative online reviews may persuade an analyst of imminent
business failure (success); the disinformation serving to sway opinion.

This issue is particularly pressing as current marketing research models
for analyzing UGC (e.g.,
\protect\hyperlink{ref-buschken2020improving}{Büschken and Allenby
2020}, \protect\hyperlink{ref-liu2020visual}{Liu et al. 2020},
\protect\hyperlink{ref-zhang2022uncovering}{Zhang et al. 2022}) operate
under the assumption that content is authentic. Therefore, they lack the
algorithms to identify and filter out inauthentic content. Employing
such models on compromised UGC data risks distorted insights---a
phenomenon we identify and demonstrate in our empirical
analysis---leading to systematic inaccuracies. This issue, while already
present, is poised to expand significantly given AI's cost-efficiencies
and capabilities.

We introduce a novel three-component methodology aimed at: (1) assessing
AI's ability to create disinformative UGC, (2) evaluating the
vulnerability of marketing analytics frameworks to such content, and (3)
exploring detection techniques through computational and statistical
models. While our primary focus is on price perception to highlight the
susceptibility of marketing research to disinformative UGC, our
methodology's adaptability enables its application across a variety of
product or service attributes. Our investigation yields the following
interconnected insights.

The AI's fabrications exhibit a wide variance across different product
and service categories. For example, a restaurant review altered by AI
might challenge the value offered by the dining experience, while a
manipulated furniture review might question an item's durability.
Furthermore, even minimal disinformative content can drastically alter
marketing model outcomes, especially since authentic UGC rarely includes
direct commentary on attributes like price. Hence, protecting marketing
research requires highly sensitive detection algorithms capable of
identifying and eliminating all forms of disinformative UGC. However,
current models, both standard ones derived from literature and those
employed by practitioners, fall short. They often allow significant
volumes of disinformative content to bypass filters, likely due to the
diversity of manipulation tactics and a reliance on identifying common
markers of fabricated content.

Our findings reveal a major vulnerability that underscores the urgent
need for regulatory and industry intervention to address AI-generated
marketing disinformation. Recent government initiatives, such as the
FTC's proposed penalties targeting fake reviews, along with grassroots
efforts like the Coalition for Trusted Reviews
(\protect\hyperlink{ref-akesson2023impact}{Akesson et al. 2023}),
reflect intensifying concerns. However, despite substantial actions
already taken by platforms, such as Amazon and Google, which manually
blocked millions of reviews in 2022
(\protect\hyperlink{ref-Knutsson_2023}{Knutsson 2023}), current measures
may not be sufficient to keep pace with rapidly evolving technologies.
Our analysis suggests that the volume of disinformative yet
realistic-seeming content could soon overwhelm manual review
capabilities.

Our paper proceeds as follows: Next, we review the theoretical
literature on UGC, generative AI, and market signal interference,
framing our key contributions. We outline our methodological approach
and framework. An in-depth discussion of our study design and methods
follows. We detail our results and provide an analysis of the findings
and their implications. The final section concludes.

\hypertarget{literature-review}{%
\subsection{Literature Review}\label{literature-review}}

Our paper contributes to several rapidly growing bodies of literature.
Given the breadth of these fields, we focus on elements of particular
relevance, with additional references available within these works.

Harnessing organic consumer perspectives, reviews, commentary, and
engagement across platforms, UGC offers uniquely rich insights into
dynamic shifts in customer sentiment, perceptions, preferences, and
trends (\protect\hyperlink{ref-liu2012survey}{Liu and Zhang 2012},
\protect\hyperlink{ref-moe2011value}{Moe and Trusov 2011},
\protect\hyperlink{ref-simonson2014marketers}{Simonson and Rosen 2014}).
The authenticity and unfiltered nature of these insights often surpass
those obtained from traditional surveys or focus groups. They accurately
gauge actual opinions and overcome respondent biases, delivering more
accurate signals for data-driven strategy and decision-making
(\protect\hyperlink{ref-ghose2019}{Ghose et al. 2019}). Consequently,
UGC has emerged as an invaluable asset, now serving as a cornerstone of
modern marketing research.

Given the vast scale and intricate nature of UGC, quantitative marketing
models and machine learning techniques have become indispensable for
extracting actionable insights from data beyond manual review.
Techniques such as natural language processing, computer vision, and
predictive modeling distill signals from the unstructured content of
customer journeys, yielding multidimensional insights
(\protect\hyperlink{ref-sebastiani2002machine}{Sebastiani 2002},
\protect\hyperlink{ref-shapiro2022measuring}{Shapiro et al. 2022}).
However, these models have historically operated without accounting for
deliberate disinformation. Furthermore, their primary focus has been on
the impact of UGC on consumer decision-making
(\protect\hyperlink{ref-mayzlin2014promotional}{Mayzlin et al. 2014}),
which has traditionally been the central concern of the UGC literature.

Diverging from prevailing focuses, our paper shifts the analytical lens
towards the potential use of marketing disinformation to distort rivals'
decision-making processes. The theoretical foundation of our research is
rooted in game theory literature on signal jamming. This framework
analyzes how information can be strategically manipulated to distort
competitors' strategic calculus. Signal jamming refers to the deliberate
introduction of misleading information into a communication channel,
thereby interfering with competitors' decision-making processes and
altering the perception of market signals
(\protect\hyperlink{ref-fudenberg1986signal}{Fudenberg and Tirole
1986}). These concepts have been applied across diverse contexts,
including corporate strategy and financial markets, as a mechanism for
obscuring information signals
(\protect\hyperlink{ref-holmstrom1999managerial}{Holmström 1999},
\protect\hyperlink{ref-zwiebel1995corporate}{Zwiebel 1995}). We adapt
this theoretical perspective to empirically examine the potential for
AI-mediated UGC to act as a form of signal jamming, disrupting marketing
research.

The focus on AI-facilitated disinformation is particularly timely, given
the current renaissance in AI and machine learning, led by advanced
language models like GPT-4. These models have achieved significant
advancements in generating content that closely mimics realistic human
writing (\protect\hyperlink{ref-brown2020language}{Brown et al. 2020}).
Notably, the ability of AI to produce content with a wide range of
stylistic nuances marks a significant departure from the more uniform
outputs of earlier automated systems. This advancement in language
modeling facilitates the seamless integration of AI-generated content
into streams of human-generated UGC. Consequently, marketing research
has begun to explore the potential of utilizing generative AI for
marketing communications, including the creation of automated
advertising copy
(\protect\hyperlink{ref-davenport2018artificial}{Davenport et al.
2018}).

However, the unregulated and diverse nature of UGC also creates fertile
ground for the proliferation of strategically crafted misleading content
(\protect\hyperlink{ref-guo2023close}{Guo et al. 2023},
\protect\hyperlink{ref-mitrovic2023chatgpt}{Mitrović et al. 2023}). The
complexities introduced by the varied styles and platforms of genuine
content serve as a form of `natural camouflage,' allowing disinformative
content to evade detection by traditional statistical pattern-based
methods (\protect\hyperlink{ref-page2012linguistics}{Page 2012}). As a
result, existing detection frameworks, including outlier analysis,
textual similarity assessments, and machine learning algorithms based on
historical patterns of fraud, face challenges in detecting fake reviews
(\protect\hyperlink{ref-perez2017automatic}{Pérez-Rosas et al. 2017},
\protect\hyperlink{ref-zhou2018fake}{Zhou and Zafarani 2018}). This
challenge is further exacerbated when the disinformative content
originates from advanced AI, given the evolving complexity and
sophistication of AI-generated disinformative content
(\protect\hyperlink{ref-crothers2023machine}{Crothers et al. 2023},
\protect\hyperlink{ref-tang2023science}{Tang et al. 2023}). This
challenge of detecting sophisticated AI-generated disinformation
underscores the urgent need for detection algorithms that offer both
heightened accuracy and computational efficiency.

\hypertarget{contributions}{%
\subsection{Contributions}\label{contributions}}

Our study introduces a novel framework for analyzing AI-fabricated UGC,
alongside strategies for its detection and mitigation on platforms such
as Amazon and Yelp. This paper innovates by incorporating in-silico
experimental manipulation in the study of disinformation. By capturing
real-world data and manipulating it using AI, we construct stimuli that
enable the development of novel statistical models and the exploration
of counterfactual scenarios. Here, we examine `what-if' questions before
their implementation in practice.

We begin by generating disinformative UGC by transforming authentic UGC.
These are then used to both train and test disinformation detection
models and to investigate real-world scenarios where UGC may be injected
into portals and bypass filters undetected. This approach addresses
several key limitations associated with relying on pre-existing UGC
data, as is typical in prior research. Most importantly, it facilitates
the study of and response to emergent threats from cutting-edge AI
technologies that may not yet be prevalent in existing observational
datasets but have the potential to be pivotal in the near future.
Moreover, this approach balances realism and experimental control by
incorporating real-world examples while allowing for systematic
adjustments.

Our findings issue a cautionary note. We discover that: (1) even minor
manipulations by AI can significantly disrupt the analytical models used
for interpreting UGC; (2) AI technologies have the capability to
seamlessly integrate manipulations into existing UGC, creating
disinformative content that is challenging to distinguish from genuine
contributions; and (3) the task of detecting these manipulations is
significantly complicated by the inherent variability in genuine UGC.
Although pooling information across reviews offers certain advantages,
it falls short of achieving the level of sensitivity required to
effectively filter out a substantial amount of disinformative UGC,
thereby preventing disruptions in inference and distortions in
decision-making processes. This underscores the urgent need for more
rigorous validation and filtration mechanisms at the point of UGC
submission, potentially involving stricter user authentication processes
and the provision of additional evidence of authenticity.

These findings underscore the exigency of addressing AI-fabricated
disinformation in UGC. The economic incentives for malicious entities to
disseminate fraudulent UGC, combined with the widespread availability of
public AI systems capable of producing such content, clearly pose a
formidable challenge. Furthermore, the rapid advancement of AI
technologies suggests the potential emergence of even more
sophisticated, non-public AI tools, which could enable the creation of
disinformative content that is increasingly difficult to detect. Given
these developments, the future landscape may be as concerning, if not
more so, than what our current observations indicate.

More broadly, our research contributes to both scholarly and practical
discussions surrounding AI-generated disinformation by highlighting
risks and presenting countermeasures. An emerging body of literature
examines the presence of AI-mediated disinformation in news
(\protect\hyperlink{ref-vosoughi2018spread}{Vosoughi et al. 2018}) and
politics (\protect\hyperlink{ref-farkas2019post}{Farkas and Schou
2019}). Our study distinguishes itself by exploring the risks associated
with unscrupulous firms exploiting generative AI for disinformative
marketing. We demonstrate how advanced AIs facilitate the mass
production of manipulated UGC that, despite appearing authentic, poses
risks of targeted information distortion at unprecedented scales
(\protect\hyperlink{ref-zellers2019defending}{Zellers et al. 2019}).
These issues are of critical importance as the rapid dissemination of
disinformative information in the digital space presents a multifaceted
challenge to not only business strategies but also public sentiment and
institutional credibility
(\protect\hyperlink{ref-allcott2017social}{Allcott and Gentzkow 2017},
\protect\hyperlink{ref-shu2017fake}{Shu et al. 2017}).

\hypertarget{methodology}{%
\section{Methodology}\label{methodology}}

\begin{figure}[htbp]
\centering
\includegraphics[width=0.8\textwidth]{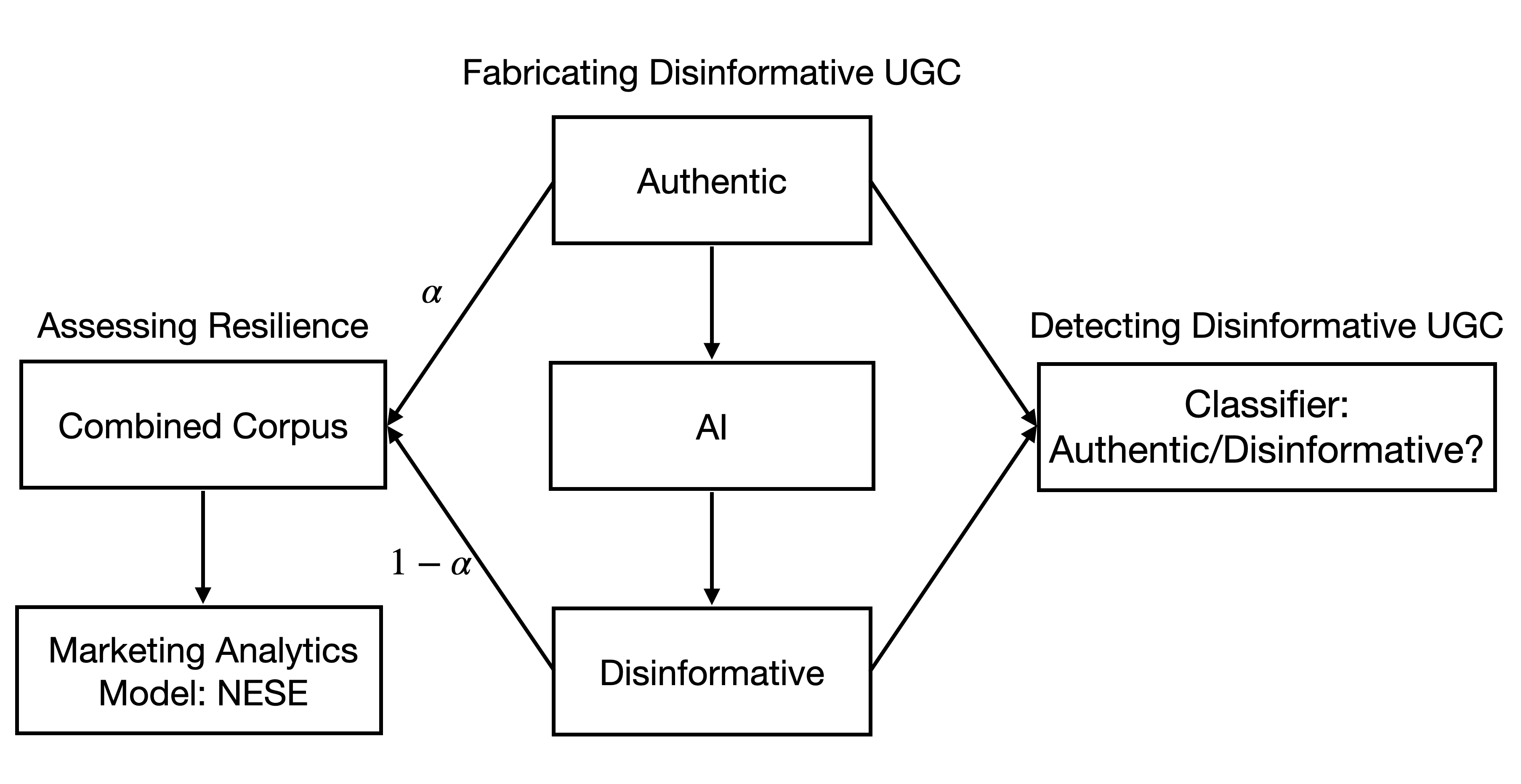}
\caption{Assessing the Impact of Disinformative UGC on Marketing research}
\begin{minipage}{\linewidth}
\medskip
\footnotesize
Note: At the center, an AI is used to convert authentic UGC into disinformative variants. To the left, authentic and disinformative UGC are blended, with the mix controlled by \( \alpha \). The blended corpus is analyzed using a marketing analytics model to evaluate the impact of disinformative UGC on marketing research. To the right, authentic and disinformative UGC are used to train a classifier to detect disinformative UGC.
\end{minipage}
\label{fig:roadmap}
\end{figure}

Our methodology is structured around a comprehensive framework designed
to assess the generation, detection, and impact of AI-fabricated
disinformation within UGC. This framework is depicted in Figure
\ref{fig:roadmap}, serving as a visual guide to our approach. At the
center of the figure, we illustrate the use of AI to generate
disinformative UGC. To the left, we detail the blending of authentic and
disinformative UGC, controlled by a parameter \(\alpha\), which varies
from 0 (completely authentic) to 1 (completely disinformative). This
mixture is analyzed using a marketing analytics model, comparing the
results from the mixed corpus to those from a purely authentic UGC
corpus, thereby quantifying the impact of disinformative content
infiltration on marketing research. To the right, we outline the use of
both authentic and disinformative UGC to train a classifier to determine
the authenticity of UGC.

The first component of our framework focuses on generating
disinformative UGC. This serves two primary purposes: it provides a
quantitative platform for exploring the distortions introduced by
disinformative content into consumer ecosystems like Amazon, and it
generates training data for developing and testing AI-fabricated UGC
detection algorithms. This component employs AI to transform original
UGC into disinformative variants. For example, the AI can generate a
review that falsely criticizes a product's pricing, even if the original
UGC was neutral or positive. Unlike existing research that either uses
platform-flagged suspicious UGC or crowdsources artificial UGC from
human assistants (\protect\hyperlink{ref-luca2016fake}{Luca and Zervas
2016}, \protect\hyperlink{ref-ott2011finding}{Ott et al. 2011}), our
approach leverages the sophisticated semantic generation capabilities of
AI and the inherent authenticity of human-authored reviews. This results
in fabricated UGC that presents increasing difficulty for detection,
complicating the formulation of countermeasures.

The second component examines the effects of disinformative UGC on
marketing analytics (\protect\hyperlink{ref-goes2014popularity}{Goes et
al. 2014}, \protect\hyperlink{ref-mohawesh2021fake}{Mohawesh et al.
2021}). Moving beyond analyses focused on consumer purchase behavior, we
propose that disinformative UGC distorts the data streams that inform
marketing research, potentially leading to strategic information
manipulation. For example, fabricated negative commentary on pricing
might drive competitors to implement ill-advised price cuts, eroding
margins and profitability. Such tactics could strategically force
competitors to underinvest or withdraw from profitable markets,
reflecting strategies akin to predatory pricing
(\protect\hyperlink{ref-mcgee1980predatory}{McGee 1980}). Therefore, to
assess the impact of disinformative content, we incrementally blend
disinformative UGC with authentic UGC, simulating varying degrees of
disinformation infiltration, and assess their implications on inference.

Our analysis employs Named Entity Sentiment Evaluation (NESE), a widely
used analytical tool (\protect\hyperlink{ref-li2017learning}{Li and Lu
2017}, \protect\hyperlink{ref-nazir2020issues}{Nazir et al. 2020}). NESE
serves two primary functions: first, it extracts named entities---the
focal subjects---from the text corpus; second, it assesses the sentiment
valences linked to these entities. As such, these correspond to the idea
of measuring the frequency with which a facet of marketing is mentioned
in UGC and the sentiment that is expressed when it is mentioned.

The term NESE is not common in marketing literature or marketing
practice, even as its application is the cornerstone of marketing
analytics. In particular, industry leaders in social media analytics
such as Awario, Mention, Talkwalker, Brandwatch, BuzzSumo, HootSuite,
and Social Mention, sell `engagement dashboards'---products that
collect, collate, and analyze UGC to derive key performance indicators
and metrics, presented visually to simplify analysis. In such products,
the two fundamental metrics are `mentions' and `sentiment' (or
trademarked brand name equivalents), with the first being the detection
of entities in UGC (e.g., brand name, price, etc.) and the second being
a detection of overall sentiment or sentiment towards each of the
detected entities---which are the precise definitions of outputs from a
NESE model, albeit termed differently. Social media dashboards are
ubiquitous; while specific data on industry size and use is difficult to
obtain and verify, to the best of our knowledge, almost all leading
brands subscribe to some form of social media analysis, which almost
always include mentions and sentiment analysis.

Existing implementations of social media analysis are either proprietary
and zealously guarded, or only designed for specific UGC streams (such
as if the implementation was developed for Twitter data). Open source
and academic implementations of NESE that serve the same purpose in
principle typically employ general-purpose labels such as `Person,'
which are often too generic or irrelevant for effective UGC analysis.
Therefore, we develop and validate a novel NESE method tailored to UGC
analysis. We apply this model to the blended data and analyze the named
entities and sentiment values recovered across various scenarios. Thus,
we ascertain how the presence of disinformative UGC alters marketing
insights, quantifying the influence of marketing disinformation.

The third component of our framework investigates the detection of
AI-fabricated UGC. We assess three distinct approaches for detecting
AI-fabricated UGC, encompassing both commercial solutions and
established machine learning techniques. The first approach involves an
evaluation of GPTZero, a leading-edge commercial tool designed to
identify AI-generated text. The second and third approaches leverage
topic intensities from topic models trained on disinformative and
authentic data (\protect\hyperlink{ref-toubia2019extracting}{Toubia et
al. 2019}). The second approach utilizes the classic Latent Dirichlet
Allocation (LDA, \protect\hyperlink{ref-blei2003latent}{Blei et al.
2003}) topic model, widely used in both academia and practice. The third
approach employs BERTopic
(\protect\hyperlink{ref-grootendorst2022bertopic}{Grootendorst 2022}), a
recent innovation in topic models that organizes embeddings into
coherent topics by synthesizing transformer-based models with
traditional TF-IDF techniques to enhance interpretability (see Table 5
in \protect\hyperlink{ref-egger2022topic}{Egger and Yu 2022}). For both
approaches, we use the topic distributions generated by separate topic
models---trained on authentic and disinformative UGC---as input features
for classifying UGC.

\hypertarget{component-1-fabricating-disinformative-ugc}{%
\subsection{Component 1: Fabricating disinformative
UGC}\label{component-1-fabricating-disinformative-ugc}}

Disinformative UGC is covertly introduced into platforms by malicious
actors to blend seamlessly with other UGC, subtly influencing consumer
opinion without detection by the platform or its users. Consequently,
the study of disinformative UGC poses a unique challenge: the data lacks
labels to distinguish between authentic and disinformative content. This
absence of labels complicates the use of standard supervised learning
methods, which rely on pre-classified data for model training, and
hinders measurement of the implications of disinformation. Previous
studies have addressed this challenge by adopting one of three
approaches: (1) leveraging data marked as suspicious by consumer
platforms such as Yelp (\protect\hyperlink{ref-luca2016fake}{Luca and
Zervas 2016}); (2) analyzing entities, specifically sellers, known to
commission disinformative UGC
(\protect\hyperlink{ref-he2022detecting}{He et al. 2022}); and (3)
crowdsourcing the creation of disinformative reviews
(\protect\hyperlink{ref-ott2011finding}{Ott et al. 2011},
\protect\hyperlink{ref-shojaee2015framework}{Shojaee et al. 2015}).
Below, we discuss each.

The first approach exhibits limitations relative to our research
question. While AI technologies have rapidly advanced, their specific
application in generating disinformative UGC has not yet become
widespread in the marketplace. Consequently, existing datasets of
platform-flagged content are much more likely to consist of
human-generated disinformation as opposed to AI-generated
disinformation. Moreover, this approach is reactive, as it is limited to
already-flagged content; it does not enable the investigation of an AI's
efficacy in generating disinformative content.

The second approach, while better aligned with our research objectives,
still falls short. It does not provide a systematic means to examine the
creation of disinformative content using AIs, nor does it offer labeled
datasets for the development and validation of detection systems or for
examining the impact of disinformative content on marketing analytics
frameworks. Therefore, it lacks the necessary depth and breadth to fully
explore the intricacies of generating, identifying, and mitigating
AI-fabricated UGC.

The third approach shares some alignment with our research goals.
However, its reliance on human-generated disinformation diverges from
our focus on AI-generated disinformation. In particular, our paper
focuses on how the data features of disinformative UGC may facilitate
its detection. For instance, studies suggest AI can exhibit stylistic
inconsistencies (\protect\hyperlink{ref-badjatiya2017deep}{Badjatiya et
al. 2017}). Additionally, while grammatically correct, AI-generated
content may suffer from contextual mismatches, and may not replicate the
emotional depth of genuine reviews
(\protect\hyperlink{ref-mukherjee2013yelp}{Mukherjee et al. 2013}).

In response, we propose a novel approach that utilizes AI to transform
the stylistic and structural templates of authentic UGC into
disinformative content. For instance, to shift consumer perception about
a product's dimensions, the AI selectively alters that aspect in the
reviews while preserving the core content. Thus, the disinformative UGC
generated does not originate `de novo' (i.e., from scratch) but results
from the AI's targeted manipulation of existing UGC.

A critical aspect of our methodology is directing AI to modify a
specific dimension of existing UGC. Although AIs lack goal-oriented
cognition, their output can be finely calibrated through tailored
prompts that provide the necessary contextual guidance. These prompts
act as algorithmic directives that embody the disinformation strategy,
instructing the AI on what to modify. For example, if the research goal
is to instill negative sentiment about product pricing, the prompt would
clearly specify this objective. Utilizing tailored prompts enables
precise calibration of the AI output to achieve predefined
disinformative objectives. This prompt-driven approach affords granular
control over the disinformative UGC generation process.

The generated data serve three interrelated objectives: (1) empirically
validating the capability of AI to generate disinformative UGC that
closely mimics authentic content, (2) rigorously evaluating
disinformation detection algorithms, and (3) assessing the systemic
impacts of disinformative UGC on consumer platforms such as Yelp and
Amazon. The framework allows for adjustments to the ratio of authentic
to disinformative UGC, enabling empirical testing across a range of
operational scenarios and enhancing real-world applicability. Our
proactive approach, unlike reactive methods constrained by retrospective
data, enables the development of countermeasures against emergent
disinformative tactics, marking a significant advancement in combating
digital disinformation. It not only enhances our understanding of
digital disinformation but it also equips practitioners with analytical
tools to stay ahead of nascent and evolving threats.

\hypertarget{component-2-measuring-the-resilience-of-inference-to-disinformation}{%
\subsection{Component 2: Measuring the Resilience of Inference to
Disinformation}\label{component-2-measuring-the-resilience-of-inference-to-disinformation}}

The second component of our framework examines the resilience of
marketing analytics models when confronted with disinformative UGC. This
exploration is crucial, as the integrity of marketing insights directly
influences strategic decision-making and, ultimately, competitive
advantage.

To systematically assess the impact of disinformative content, we blend
disinformative with authentic data, simulating varying degrees of
disinformation infiltration. We utilize NESE, adapted to focus on
entities and sentiments pivotal in marketing analysis, such as product
quality, brand perception, and customer satisfaction. We create blended
datasets where the proportion of disinformative content ranges from 0\%
(completely authentic) to 100\% (completely disinformative), increasing
in 1\% increments. We apply NESE to extract and evaluate named entities
and their associated sentiments in each dataset. This allows us to
observe the shift in marketing insights as the level of disinformation
increases, providing a clear picture of the implications of
disinformative UGC.

Consider, for example, a scenario where a product is generally praised
for its durability in genuine UGC. As disinformative UGC criticizing the
product's durability is introduced, NESE might begin to register
`durability' more frequently as a negative entity, despite its positive
portrayal in authentic reviews. This shift could mislead firms into
believing there is a widespread issue with their product, potentially
prompting unnecessary strategic pivots.

The application of NESE to UGC data presents a unique challenge.
Typically, NESE models are designed to identify and evaluate standard
entity classes such as `Person' (e.g., `John Smith'), `Organization'
(e.g., `United Nations'), and `Location' (e.g., `Paris'), and are
generally trained on broad, unstructured text data. This training often
results in generic labels that lack the specificity needed for effective
and detailed UGC analysis.

To address this challenge, we develop a novel method for conducting NESE
analysis using AI. We present the AI with a prompt that describes the
context in which a review was written (e.g., on Amazon), specifies the
entities of interest, defines the sentiment analysis scale, and outlines
the desired output format. This method allows analysts to specify
relevant entities and sentiment scales, generating tailored algorithmic
outputs for their research. Simultaneously, it leverages AI's robust
capacity for content analysis, facilitating the detection of specified
entities even when they are subtly mentioned, and accurately discerning
the sentiment associated with them. This offers a particular advantage
over traditional statistical models, which often struggle with
linguistic complexities like polysemy, synonymy, and long-range
dependencies---challenges that modern AI is adept at handling.

We develop the prompt as follows: We instruct an AI to examine all
original UGC to identify any product-related attributes potentially
relevant to a firm, such as product quality, size, and pricing. The
objective is to isolate key entities in each piece of UGC that hold
relevance for the analyst. For instance, if a piece of UGC discusses
aspects such as product quality, size, and pricing, these entities are
to be identified. Next, we compile the identified entities, focusing on
the most salient ones based on frequency and perceived significance.
This refines our list to a focused set of key entities. We use this list
to create a final prompt that directs the AI to first identify the
presence of these selected entities in the UGC, and then to assess the
sentiment towards each, outputting the results in a computationally
amenable format.

\hypertarget{component-3-detecting-disinformation}{%
\subsection{Component 3: Detecting
Disinformation}\label{component-3-detecting-disinformation}}

The third component of our framework studies the detection of
AI-fabricated UGC. We assess three distinct approaches for detecting
AI-fabricated UGC, encompassing both commercial solutions and
established machine learning techniques. The first approach involves an
evaluation of GPTZero, a leading-edge commercial tool designed to
identify AI-generated text. It operates on a simple yet effective
principle: the tool contains an internal AI that mimics the behavior of
popular generative models such as GPT-4 and Llama. When GPTZero analyzes
a piece of text, it masks parts of the content and feeds the remainder
into its internal AI. This internal AI predicts the likelihood of it
generating the masked content. A high likelihood suggests the content
matches patterns typical of AI-generated content, indicating it is
AI-generated. Human-written text tends to diverge in linguistic patterns
from the internal AI, resulting in a lower likelihood. GPTZero leverages
this differential to categorize texts as either AI-generated or
human-authored.

For effective differentiation between authentic and disinformative
content, it is essential to furnish a classifier with informative
features that distinctly highlight the key differences between these two
types of UGC. To this end, the second and third models are based on
topic intensities. Specifically, we train two distinct topic
models---one on authentic UGC and the other on disinformative UGC.

The topic distributions generated by these models encapsulate the unique
lexical and semantic patterns characteristic of each type of content. We
incorporate the topic distributions as features
(\protect\hyperlink{ref-toubia2019extracting}{Toubia et al. 2019}) in a
supervised feedforward network classifier, which is trained to detect
AI-generated disinformative UGC. This enables the classifier to be
well-equipped to discern whether a document aligns more closely with
patterns observed in either the authentic or disinformative topic model;
the relative intensity of topics from both models provides a robust
statistical basis for categorizing UGC documents as either genuine or
artificially fabricated. The classifier acts as a universal
approximator, utilizing the discriminative features encoded in the topic
distributions to discern semantic and stylistic patterns indicative of a
content's origin.

The second model relies on LDA, a classic topic modeling technique,
widely used in both academia and practice
(\protect\hyperlink{ref-berger2023wisdom}{Berger and Packard 2023}).
However, LDA has notable limitations. As a probabilistic model fitted
using likelihood-based inference methods, LDA employs a simplified
mixing process where each token is generated independently of prior
tokens. This assumption does not account for multi-word phrases and the
dependency of word meaning on prior word uses (e.g., the difference
between the meaning of `bank' in `The fisherman walked along the bank of
the river' and `She went to the bank to deposit her paycheck'). Such
nuances cannot be accommodated in LDA. Additionally, LDA requires
frequent word use in the training dataset, as word meaning is inferred
implicitly from the training data.

The third model harnesses the capabilities of BERTopic, a prominent
neural topic modeling technique in Natural Language Processing. BERTopic
is based on recent advancements involving the use of generative
pretrained models such as BERT, which facilitates analysis through
transfer learning, whereby common uses of words are encoded in numerical
representations. BERTopic uses these numerical representations to form
document representations that are then used to infer topics and topic
intensities. We further augment standard BERTopic by incorporating
embeddings from GPT-4, a powerful language model that encodes semantic
and stylistic information. This enhancement enriches the interpretive
layers of our model, enabling it to account for nuanced word order and
phrase interactions that conventional statistical topic models like LDA
cannot capture.

\hypertarget{data}{%
\section{Data}\label{data}}

Our study utilizes two real-world UGC datasets from Amazon and Yelp.
Specifically, we accessed a dataset from Amazon covering the years 1996
to 2018, made publicly available by Ni et al.
(\protect\hyperlink{ref-ni2019justifying}{2019}). This dataset was
selected for its broad acceptance as a reliable source in academic
research and its accessibility. Our analysis focused on the Electronics
category, representing differentiated products. To complement this
product-focused analysis and broaden its applicability, we also included
a dataset from Yelp, which consists of service reviews.

The original datasets comprise 20,994,353 reviews from Amazon and
6,990,280 reviews from Yelp, spanning 11 major metropolitan areas. To
enable a robust analysis while ensuring feasibility, we curated a sample
of reviews strategically chosen to meet two key objectives: 1) to
provide sufficient statistical power for valid inference, and 2) to
maintain the dataset at a manageable size to facilitate replication and
extension by other researchers.

We conducted a power analysis to determine the sample size. While
examining the effects of disinformative UGC on marketing insights
requires a relatively modest number of samples, the development,
training, and validation of a disinformation detection algorithm
necessitate substantially more data points. Preliminary testing
indicated that an optimal configuration for our model would include
approximately 10 topics for the authentic UGC corpus. Assuming a similar
10-topic model for the disinformative UGC, we estimated that
approximately 20 topic distributions would serve as features for these
models. Given a feedforward network with 20 input features, 16 hidden
units, and 1 output unit, which has 336 trainable parameters, and
following the rule of thumb of 10 observations per parameter---doubling
this number for robustness---we estimated a requirement of around 8,000
observations to train the model effectively. Therefore, to accommodate
our planned 80-10-10 percentage split of training, validation, and test
data, we aimed for a final sample size of 10,000 reviews from each
dataset.

Several preprocessing steps were implemented to enhance the quality and
consistency of the selected data. Initially, reviews containing
non-English characters were removed to minimize potential parsing issues
and align with our use of an English language AI. Next, we excluded
extremely short or long reviews, defined as those in the lowest and
highest quartiles by length, to mitigate outlier influence. Duplicate
reviews were also eliminated to prevent overrepresentation. This
filtering process yielded clean, high-quality data suitable for rigorous
analysis. We then randomly selected the final sample of 20,000
reviews---10,000 from each dataset---from this filtered data. During our
analysis, one observation in the Amazon sample was identified as
containing a malformed review. To maintain balance between the datasets,
this review and a randomly selected review from the Yelp sample were
excluded. Consequently, the results presented in this study are based on
analyses of 9,999 reviews from each dataset.

The dates on which the reviews were posted predate the emergence of
modern AIs, indicating that all reviews were written by humans and not
generated by AI. Thus, we designate these reviews as `authentic' and use
`disinformative' for AI-fabricated reviews. Furthermore, since
AI-fabricated reviews are most likely intended for fraudulent purposes,
we use the terms `AI-fabricated' and `disinformative' interchangeably in
our analysis. Future applications of AI could potentially generate
disinformative authentic reviews, for example, by more directly
interacting with the user or observing behavior before generating a
review. However, such reviews are beyond the scope of our current
research, which focuses on reviews that are strategically
disinformative.

\begin{figure}[htbp]
\centering
\begin{subfigure}{.5\textwidth}
  \centering
  \includegraphics[width=\linewidth]{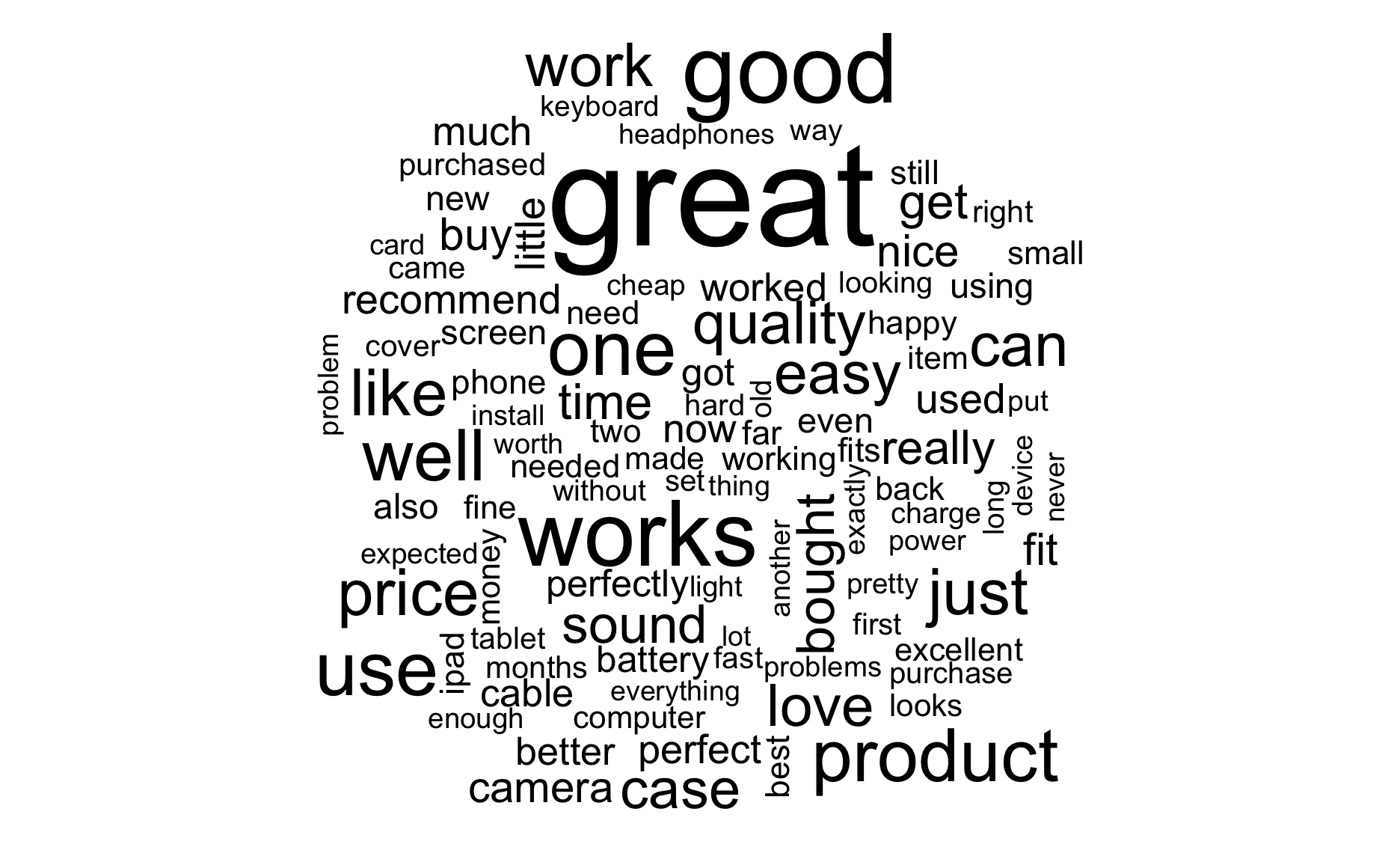}
  \caption{Amazon Data}
\end{subfigure}%
\begin{subfigure}{.5\textwidth}
  \centering
  \includegraphics[width=\linewidth]{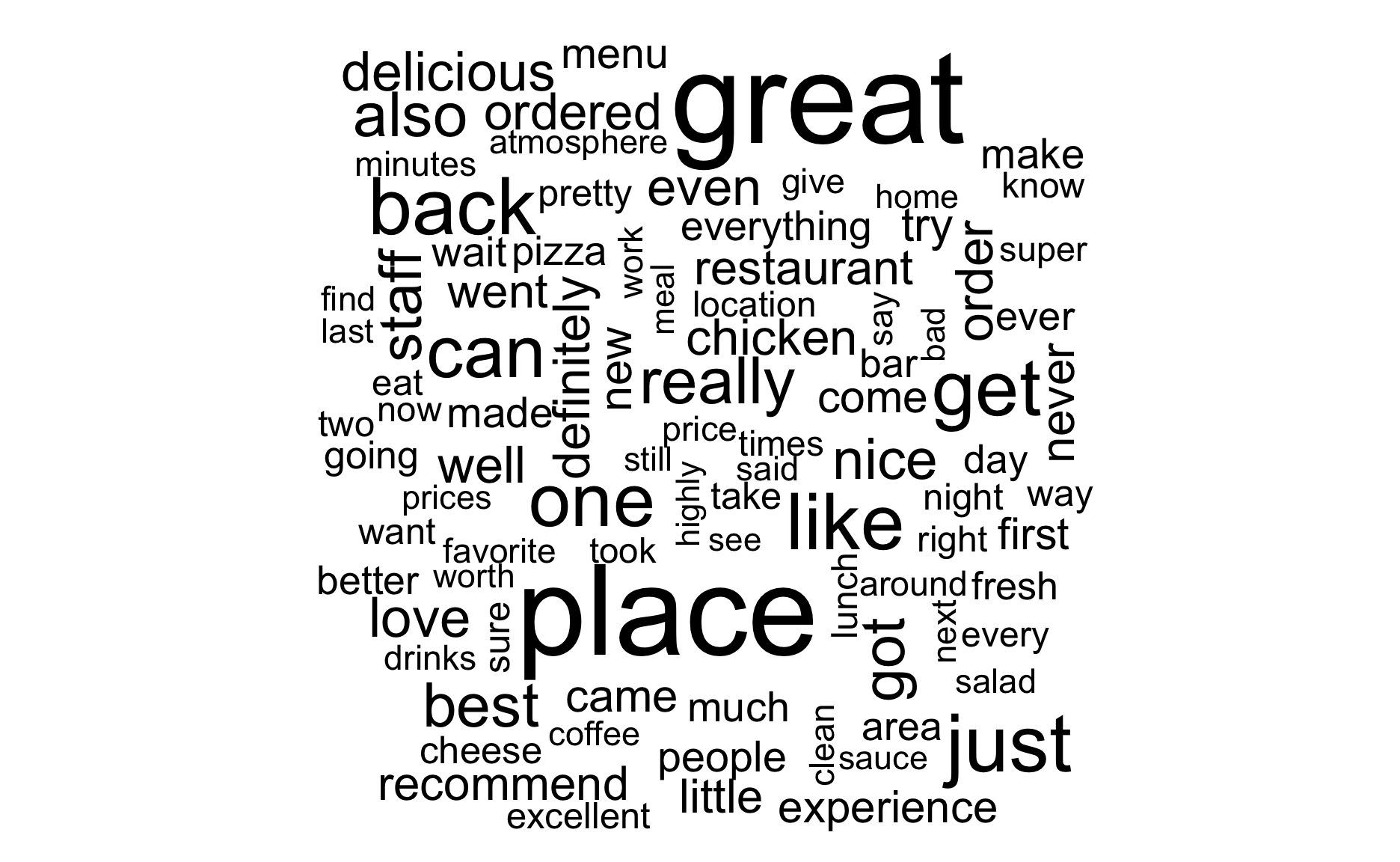}
  \caption{Yelp Data}
\end{subfigure}
\caption{Word Clouds of Amazon and Yelp Data}
\label{fig:wordclouds}
\begin{minipage}{\linewidth}
\medskip
\footnotesize
Note: The word clouds are derived from both Amazon and Yelp data after excluding common stopwords and terms that do not contribute to thematic analysis.
\end{minipage}
\end{figure}

A thematic analysis of UGC from Amazon and Yelp, as illustrated in the
word clouds in Figure \ref{fig:wordclouds}, reveals distinct consumer
priorities in product and service domains. On Amazon, the high frequency
of evaluative terms like `great' (2808 instances) and `good' (1769),
alongside functional descriptors such as `product' (1252), `price'
(1062), and `quality' (939), suggests a lexicon focused on utilitarian
assessments. The emphasis on specific attributes like `sound' and
`camera' further demonstrates a granular focus on product functionality,
consistent with multi-attribute utility models.

In contrast, Yelp reviews reflect the experiential nature of service
consumption. Terms like `food', `service', and `delicious' highlight an
experiential evaluation framework beyond just objective quality. The
emphasis on satisfaction and affective responses aligns with literature
on experiential consumption (e.g.,
\protect\hyperlink{ref-holbrook1982experiential}{Holbrook and Hirschman
1982}). The presence of terms like `friendly' and `atmosphere' connects
to research on ambient factors influencing customer experiences. Both
datasets prominently feature sentiment-laden words, albeit with Yelp's
lexicon skewing towards more emotive expressions like `love' (Yelp: 1433
vs.~Amazon: 908) and `delicious' (Yelp: 1410). Together, these lexical
patterns demonstrate a social, atmospheric focus aligning with the
experiential nature of services (e.g.,
\protect\hyperlink{ref-bitner1992servicescapes}{Bitner 1992}).
Crucially, the observed thematic patterns not only reveal inherent
contrasts between product and service reviews but also showcase the
multidimensional nature of modern consumer feedback: Amazon reviews tend
to be informative and feature-centric, while Yelp content offers richer,
more expressive narrations of experiences.

\hypertarget{results}{%
\section{Results}\label{results}}

Price perception is a critical factor often emphasized in genuine UGC,
where products may be criticized as overpriced in relation to their
quality or size. In this study, we explore a scenario in which a
competitor seeks to undermine rivals' marketing research by introducing
disinformation into UGC, suggesting consumer dissatisfaction with
prices. The intent is to coerce rivals into unsustainable price
reductions, thereby diminishing their profits and investment potential.
This scenario illustrates our focus on unfair competition through
disinformative UGC. However, the applicability of our framework extends
beyond this instance. For instance, it is equally relevant to situations
where firms might deploy disinformative UGC to insinuate that a
competitor's prices are too low, potentially facilitating price
inflation through collusive pricing strategies.

\hypertarget{component-1-fabricating-disinformative-ugc-1}{%
\subsection{Component 1: Fabricating disinformative
UGC}\label{component-1-fabricating-disinformative-ugc-1}}

Table \ref{table:comparison_amazon} and Table
\ref{table:comparison_yelp} illustrate our methodology with a selection
of authentic and disinformative UGC from Amazon and Yelp, chosen to
represent the diversity of transformations across product and service
categories. The left column presents the original reviews, capturing
genuine customer sentiments, while the right column presents their
disinformative counterparts, crafted to introduce skepticism about the
products' value for money. The prompt guiding the construction of the
disinformative variants is detailed below\footnote{The presented prompts
  were developed through careful pretesting, where we observed that the
  AI's performance improved when using more detailed prompts with
  repeated information. We conjecture that this occurs because the
  repeated information reduces the probability of the AI misinterpreting
  the instructions.}.

\singlespacing
\begin{quoting}
\noindent Subtly alter the following user-generated content (UGC) to introduce a nuanced negative perspective on the product's value for money.
\noindent 1. Comprehend Original Tone: Carefully analyze the UGC to understand the current sentiment, style, and any commentary on product value.
\noindent 2. Identify Target Areas: Locate key phrases specifically related to discussions of the product's value to subtly modify. Choose sections where minimal changes can shift the implied sentiment.
\noindent 3. Make Minimally Invasive Edits: Make minimal alterations to target areas, focused on discreet wording adjustments that shift value perceptions from positive/neutral to negative.
\noindent 4. Maintain Coherent Flow: Integrate changes seamlessly so edits feel native to the original text. Retain the original voice, tone, and structure.
\noindent 5. Exercise Utmost Subtlety: Edits should be virtually indistinguishable from the original. Avoid conspicuous changes while introducing the intended sentiment shift.
\noindent Execute these instructions with careful judgment and precision to preserve the genuine feel of the original text. The edits must not be readily perceptible after insertion.
\end{quoting}
\doublespacing

\begin{table}[htbp]
\centering
\caption{Comparison of Authentic and disinformative UGC--Amazon Reviews}
\label{table:comparison_amazon}
\begin{tabular}{>{\raggedright\arraybackslash}p{7.5cm}>{\raggedright\arraybackslash}p{7.5cm}}
\toprule
\textbf{Authentic UGC} & \textbf{Disinformative UGC} \\
\midrule
I am totally satisfied with this product and with the service provided. My first try I did not push the connector in fully (the first instruction for installation). Once it clicked into the iPad, my pictures loaded perfectly. The delivery service was speedy too. & While I am generally content with this product and the service provided, I did have a minor hiccup on my first try. I didn't push the connector in fully, which is the first instruction for installation. Once it clicked into the iPad, my pictures loaded, but not as swiftly as I'd hoped. The delivery service, while prompt, didn't quite match up to the price I paid.\\
\addlinespace
Love this item. Has worked great. Good quality for small price. Would order again for gifts. Nothing else to say. & Like this item. Has worked okay. Decent quality for a small price. Might consider ordering again for gifts. Nothing else to say, but the value for money could be better.\\
\bottomrule
\end{tabular}
\begin{minipage}{\linewidth}
\medskip
\footnotesize
Note: This table juxtaposes authentic and disinformative UGC.
\end{minipage}
\end{table}

\begin{table}[htbp]
\centering
\caption{Comparison of Authentic and disinformative UGC--Yelp Reviews}
\label{table:comparison_yelp}
\begin{tabular}{>{\raggedright\arraybackslash}p{7.5cm}>{\raggedright\arraybackslash}p{7.5cm}}
\toprule
\textbf{Authentic UGC} & \textbf{Disinformative UGC} \\
\midrule
By far one of the best Italian restaurants in Saint Louis. They are known for their pizza and it is good, but my love at Nick and Elena's is their salad.
The special salad comes served with mozzarella slices, pepperoni, tomato, ham, salami, pepperoncini, anchovies, and black olives. It is then covered in their AMAZING Italian dressing. I usually have them put the dressing on the side as well as the anchovies. We tend to split this dish as a family so we need it on the side, because I do not like anchovies, so keep them away please!
The best part about this place is the atmosphere. It is family-run and it really feels that way.&
While it's often touted as one of the best Italian restaurants in Saint Louis, particularly for their pizza, my personal favorite at Nick and Elena's is their salad, though it's a bit pricey for what it is.
The special salad is served with mozzarella slices, pepperoni, tomato, ham, salami, pepperoncini, anchovies, and black olives. It's then doused in their Italian dressing, which is admittedly quite good. I usually request the dressing and the anchovies on the side, most likely to justify the cost. We tend to split this dish as a family, so we need it on the side, especially since I'm not a fan of anchovies.
The most redeeming quality of this place is the atmosphere. It's family-run and it really feels that way, but it's hard to ignore the pinch on the wallet.\\
\addlinespace
The Good: The restaurant was not crowded between 5 and 6 on a Wednesday evening. The service was very good, our server was pleasant and attentive. The food was good, everyone at our table enjoyed their meal. In addition, the dessert was delicious and enough to share.
The Bad: For those of you with children you may expect crayons and an activity sheet. Well, my toddler was given an activity sheet, but they were out of crayons. Other than that, nothing else stands out as being bad.
Overall, this is a good place for a casual meal with family or friends. The food was good, but not great. I will go back again at some point in the future. Enjoy! &
The Good: The restaurant was not crowded between 5 and 6 on a Wednesday evening. The service was very good, our server was pleasant and attentive. The food was good, everyone at our table enjoyed their meal. In addition, the dessert was delicious and enough to share.
The Bad: For those of you with children you may expect crayons and an activity sheet. Well, my toddler was given an activity sheet, but they were out of crayons. Other than that, nothing else stands out as being bad.
Overall, this is a decent place for a casual meal with family or friends. The food was good, but not great, and considering the price, it didn't quite hit the mark on value for money. I give it just above 3 stars. I might go back again at some point in the future, but it's not a priority. Enjoy! \\
\bottomrule
\end{tabular}
\begin{minipage}{\linewidth}
\medskip
\footnotesize
Note: This table juxtaposes authentic and disinformative UGC.
\end{minipage}
\end{table}

Table \ref{table:comparison_amazon} showcases the AI's proficiency in
modifying Amazon reviews to introduce skepticism while preserving the
original sentiment and style. For example, a review expressing complete
satisfaction with a product (`I am totally satisfied with this product
and with the service provided\ldots{} The delivery service was speedy
too.') is altered to convey moderate contentment and question the
product's value (`While I am generally content\ldots{} didn't quite
match up to the price I paid.'). Similarly, a positive review (`Love
this item. Has worked great. Good quality for small price. Would order
again for gifts. Nothing else to say.') is transformed to inject a hint
of skepticism (`Like this item. Has worked okay\ldots{} but the value
for money could be better.'). These minimal yet strategic adjustments
introduce doubt about pricing, subtly shifting the narrative without
disrupting the original review's tone.

Table \ref{table:comparison_yelp} demonstrates similar patterns in Yelp
reviews. An authentic review praising an Italian restaurant in Saint
Louis (`By far one of the best Italian restaurants in Saint
Louis\ldots{}') is contrasted with its disinformative counterpart that
questions the value for money (`While it's often touted as one of the
best\ldots{} it's a bit pricey for what it is.'). Another review,
originally balanced in discussing a dining experience, is altered to
express skepticism about the price-value ratio (`The food was good, but
not great, and considering the price, it didn't quite hit the mark on
value for money\ldots{}'), changing a straightforward recommendation to
a more cautious endorsement.

These examples illustrate the AI's ability to modify authentic UGC into
disinformative variants, focusing on altering perceptions of product or
service value. They highlight the challenge of detecting such
disinformation, as the disinformative reviews closely mimic the original
content with minimal changes, yet achieve the overarching goal through
distinct alterations across different UGC. However, these instances also
suggest the potential for AI to detect the manipulations. The
disinformative process, when applied broadly, may lead to awkward
phrasing or out-of-place caveats that stand out in otherwise positive
reviews, such as `didn't quite match up to the price I paid' or
`considering the price, it didn't quite hit the mark on value for
money.' Identifying these incongruities, inherent to the AI-generation
process, is precisely how we hope to identify and prevent the spread of
AI-generated disinformative UGC.

\hypertarget{component-2-measuring-the-resilience-of-inference-to-disinformation-1}{%
\subsection{Component 2: Measuring the Resilience of Inference to
Disinformation}\label{component-2-measuring-the-resilience-of-inference-to-disinformation-1}}

\hypertarget{nese-model-development.}{%
\paragraph{NESE: Model Development.}\label{nese-model-development.}}

To develop a NESE model, we aimed to compile a list of key marketing
entities present in the UGC under examination. For this purpose, the
following prompt was presented to an instance of a AI alongside a single
piece of authentic UGC. To ensure data integrity and prevent information
leakage, each piece of UGC was evaluated independently by a separate
instance of the AI. This process, conducted with 19,998 distinct
instances, yielded 19,998 distinct sets of entities, with each set
originating from a separate piece of UGC.

\singlespacing
\begin{quoting}
\noindent Your task is to analyze user-generated content (UGC) from online platforms such as Amazon and Yelp, with the aim of identifying and listing key entities relevant to marketing analysis. Please adhere to the following instructions:
\noindent 1. Read the UGC Carefully: Thoroughly read the UGC to fully grasp the context and content.
\noindent 2. Identify Key Entities: Focus on identifying marketing-related entities within each UGC, such as product features, brand names, service aspects, pricing information, and customer experiences. Ensure to exhaustively list all such entities found in each UGC.
\noindent 3. List Entities Individually: For each UGC, create a list of all identified entities, such as 'Product Quality', 'Pricing', 'Brand Name'.
\noindent 4. Format Output: Structure your output as follows: \#\#\#\#\#Identified Entities: [entities here]\#\#\#\#\#, where [entities here] is a comma-separated list of all identified entities from the UGC. It's imperative to strictly adhere to this format to facilitate programmatic parsing of the data. If no clear entities are identified, output \#\#\#\#\#Identified Entities: NONE\#\#\#\#\#.
\noindent Please execute these instructions with precision.
\end{quoting}
\doublespacing

We tabulated the extracted entities and excluded those observed fewer
than five times, as these instances represented less than 0.05\% of the
data. This exclusion criterion allowed us to focus our analysis on the
more prevalent entities, enhancing the robustness and relevance of our
findings. Additionally, this approach streamlined the computational
demands associated with the filtering task.

The aggregated entity counts, now represented as entity-frequency pairs,
were presented to an AI model. This model was equipped with a context
window sufficiently large to process the entire set of entities
identified from 9,998 pieces of diverse UGC in a single interaction. By
doing so, we could integrate knowledge across the entire context window
and enumerate entities. We presented the entity-frequency data from the
two datasets separately to the AI, which allowed us to discern distinct
topics related to products and services. The AI was then tasked with
identifying the 10 most important marketing-related entities, determined
by their relevance and potential impact on marketing strategies and
decision-making.

\singlespacing
\begin{quoting}
\noindent Your task is to analyze a string of data representing marketing-related entities extracted from user-generated content (UGC) on platforms like Amazon and Yelp. The data is formatted as a series of entity-frequency pairs, where each entity is listed alongside the frequency of its occurrence, separated by commas. For example, in the string 'entity1': 5, 'entity2': 3, 'entity3': 4, 'entity1' was observed 5 times, 'entity2' 3 times, and 'entity3' 4 times. Your objective is to identify and list the entities that are most important for marketing analysis, considering both their relevance and frequency of appearance in the data. Please follow these steps:
\noindent 1. Parse the Data: Interpret and segregate the data correctly, focusing on the entity-frequency pairs. Each pair is separated by a colon and individual pairs are separated by commas.
\noindent 2. Consolidate Similar Entities: Group similar entities, accounting for variations such as singular vs. plural forms or slight spelling differences (e.g., 'Price' and 'Prices', 'BrandName' and 'Brand-Name').
\noindent 3. Assess Entity Importance: Examine the entities for their significance in marketing analysis. Prioritize not just based on frequency but also consider the potential impact each entity could have on marketing strategies and decision-making.
\noindent 4. Identify 10 Key Entities: Determine the 10 entities that are most crucial for understanding the marketing dynamics represented in the UGC, taking into account both their occurrence and strategic importance.
\noindent 5. List Important Entities: Compile a list of the 10 entities identified as most important, providing a brief explanation for each regarding why it is critical for marketing analysis. Present the entities in descending order of importance, with each entity and its explanation on a separate line.
\noindent Your thorough analysis is essential for the development of an enhanced marketing analytics model. Your ability to discern the significance of each entity and its impact on the overall marketing landscape is key to this analysis.
\end{quoting}
\doublespacing

In this study, we adopted a data-driven approach to identify entities,
ensuring a comprehensive analysis of the influence of disinformative UGC
on marketing analytics. For broader applications of our methodology and
framework, various ancillary strategies can be employed to develop
similar lists. This includes direct consultation with marketing teams,
who can identify the variables most crucial to their decision-making
processes. Such an approach would tailor the analysis to specific
contexts and questions without necessitating substantial modifications
to other components of our methodology.

\begin{table}[htbp]
\centering
\caption{Key Marketing Entities in Amazon and Yelp Datasets}
\label{tab:entities}
\begin{tabularx}{0.8\textwidth}{X|X}
\toprule
\textbf{Entities in Amazon Reviews} & \textbf{Entities in Yelp Reviews} \\
\midrule
Product Quality & Customer Experience \\
Customer Experience & Pricing \\
Pricing & Food Quality \\
Brand Name & Service Quality \\
Delivery Time & Customer Service \\
Product Functionality & Product Quality \\
Customer Service & Brand \\
Ease of Use & Atmosphere \\
Product Features & Cleanliness \\
User Experience & Staff Friendliness \\
\bottomrule
\end{tabularx}
\begin{minipage}{\linewidth}
\medskip
\footnotesize
Note: This table juxtaposes key marketing entities in descending order of frequency of appearance in the Amazon and Yelp datasets.
\end{minipage}
\end{table}

The key entities in the Amazon and Yelp datasets (presented in Table
\ref{tab:entities}) provide insightful indications of customer
priorities and perceptions across various market segments. In Amazon
reviews, the emphasis on `Product Quality' and `Product Functionality'
highlights the significance of tangible product aspects in e-commerce.
These elements are vital as customers rely on descriptions and reviews
to guide their purchasing decisions without physical product evaluation.
The prominence of `Customer Experience' and `Ease of Use' underscores
the importance placed on user-friendly interfaces in electronic
products, reflecting the product categories prevalent in Amazon reviews.
Conversely, the Yelp dataset demonstrates a more pronounced focus on
experiential and service-related factors. `Customer Experience' and
`Service Quality' emerge as key entities, mirroring Yelp's focus on
service-based businesses such as restaurants and hospitality. The
importance of `Food Quality' and `Atmosphere' is also notable, affirming
their influence on customer perceptions within these sectors.

The prominence of `Pricing' as a key entity in both datasets underscores
its critical, overarching role across various platforms and industries.
Its influence extends beyond immediate purchasing decisions to impact
broader market positioning and consumer value perceptions. This finding
aligns with our study's emphasis on the manipulation of price
perceptions through fraudulent UGC, underscoring its role as a
significant factor in information disruption. Conversely, the focus on
entities like `Delivery Time' in Amazon and `Cleanliness' in Yelp
underscores the distinct operational priorities of e-commerce and
service-oriented businesses. Amazon's prioritization of efficient
logistics contrasts with Yelp's emphasis on the tangible aspects of
service environments. Lastly, five keywords---`Customer Experience',
`Pricing', `Product Quality', `Customer Service', and `Brand'---overlap
between both sets of reviews, reflecting universal elements of the
marketing mix that resonate with consumers whether they are shopping
online or visiting a service-oriented establishment.

Five entities were present in both the Amazon and Yelp lists; therefore,
the final list comprises 15 unique marketing entities. We formulated a
prompt that encompasses these 15 entities, instructing the AI to: (1)
determine whether each entity is mentioned in the UGC, and (2) assess
the sentiment expressed towards the entity, using a scale from 1 to 7.
Here, 1 indicates strong negative sentiment, and 7 indicates strong
positive sentiment. The output of the AI mirrors that of a traditional
NESE model, which identifies entities and gauges the polarity of
sentiments expressed towards them.

\singlespacing
\begin{quoting}
\noindent Your task is to analyze user-generated content (UGC) from online platforms such as Amazon and Yelp, with the aim of named entity recognition and sentiment evaluation. Please adhere to the following instructions:
\noindent 1. Read the UGC Carefully: Thoroughly read the UGC to fully understand the context and content.
\noindent 2. Named Entity Recognition: Identify if any of the following 15 marketing entities are mentioned in the UGC: Product Quality, Customer Experience, Pricing, Brand Name, Delivery Time, Product Functionality, Customer Service, Ease of Use, Product Features, User Experience, Food Quality, Service Quality, Atmosphere, Cleanliness, Staff Friendliness. Concentrate solely on these entities.
\noindent 3. Sentiment Evaluation: If an entity is mentioned, evaluate the polarity of the sentiment expressed towards it on a scale from 1 to 7. Use '1' to signify strong negative sentiment, '4' for neutral sentiment, and '7' for strong positive sentiment. Intermediate values (2-3 and 5-6) should reflect varying degrees of negative and positive sentiments, respectively. Ensure that your sentiment assessment accurately reflects the context and tone of the UGC.
\noindent 4. Record Findings: Document each pair of recognized named entity and the corresponding sentiment score. Format your output as follows: {'Entity: [entity name], Sentiment Score: [score]'}, listing each entity-sentiment pair separately with commas between pairs.
\noindent Your precision in executing these instructions is crucial to accurately gauge the sentiment towards the identified entities.
\end{quoting}
\doublespacing

\hypertarget{disinformative-ugc-infiltration}{%
\paragraph{Disinformative UGC
Infiltration}\label{disinformative-ugc-infiltration}}

We implemented the NESE model on our dataset of 39,996 pieces of
UGC---equally divided between authentic and disinformative UGC from the
Amazon and Yelp platforms. The impact of AI-generated manipulations
becomes clear when examining the summary statistics. In the authentic
Amazon UGC, `price' was identified as an entity in 2,140 instances,
making up 21.4\% of the content. In stark contrast, in the
disinformative Amazon UGC, `price' was detected far more
frequently---7,726 times, or 77.3\% of the content. Moreover, the
average sentiment towards price in the authentic Amazon UGC was 5.86 (on
a scale from 1 to 7), indicating a predominantly positive sentiment. On
the other hand, the disinformative UGC exhibited a significantly
negative sentiment, with an average of 2.82. Turning our attention to
the Yelp data, `price' was identified in 2,753 instances of authentic
UGC, or 27.5\% of the time, with an average sentiment of 4.67. This
reveals that sentiment towards price in authentic Yelp UGC is generally
less positive compared to Amazon's data. In the disinformative Yelp UGC,
`price' was detected in 8,640 instances (86.4\%), with the average
sentiment plummeting to 2.58. These findings highlight the AI's efficacy
in altering authentic UGC to introduce a negative sentiment towards
price, thereby impacting price mentions and sentiment analysis---two
critical metrics in both academic research and industry practice.

To further explore disinformative UGC infiltration, we curated a
subsample consisting of 2,000 pieces of UGC from each dataset---half
authentic and half disinformative, totaling 4,000 pieces of UGC. This
approach simulates scenarios where analysts might encounter AI-generated
disinformative UGC that doesn't directly align with the observed
authentic UGC. For each piece of authentic UGC in the subsample, its
disinformative counterpart was excluded. By selecting only 10\% of the
authentic UGC, we retained the majority of the original sample for the
inclusion of disinformative UGC. Applying the designated prompt to this
subsample yielded 4,000 sets of results, each comprising
entity-sentiment pairs from individual pieces of UGC.

\begin{figure}[htbp]
\centering
\begin{subfigure}{.5\textwidth}
  \centering
  \includegraphics[width=\linewidth]{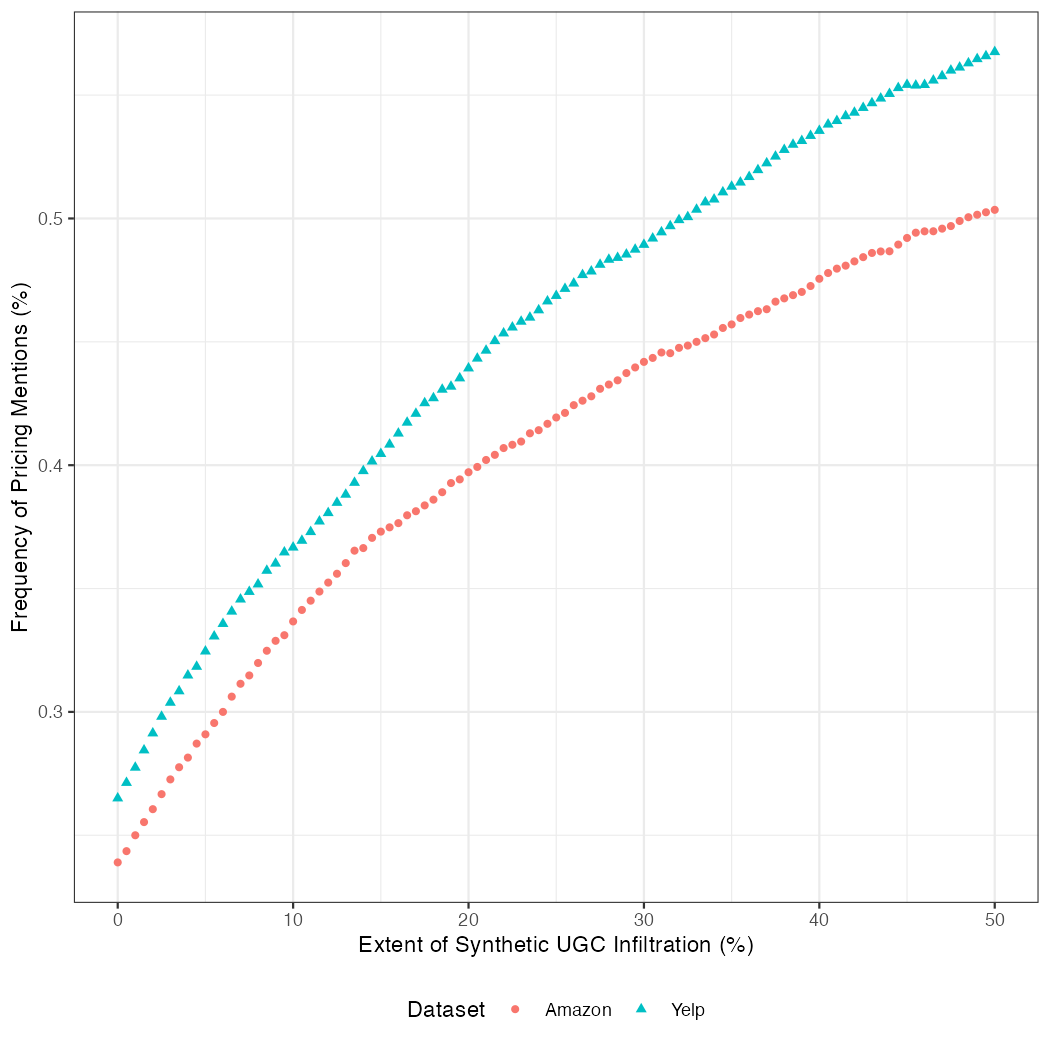}
  \caption{Frequency of Price Mentions}
  \label{fig:Price_Mentions}
\end{subfigure}%
\begin{subfigure}{.5\textwidth}
  \centering
  \includegraphics[width=\linewidth]{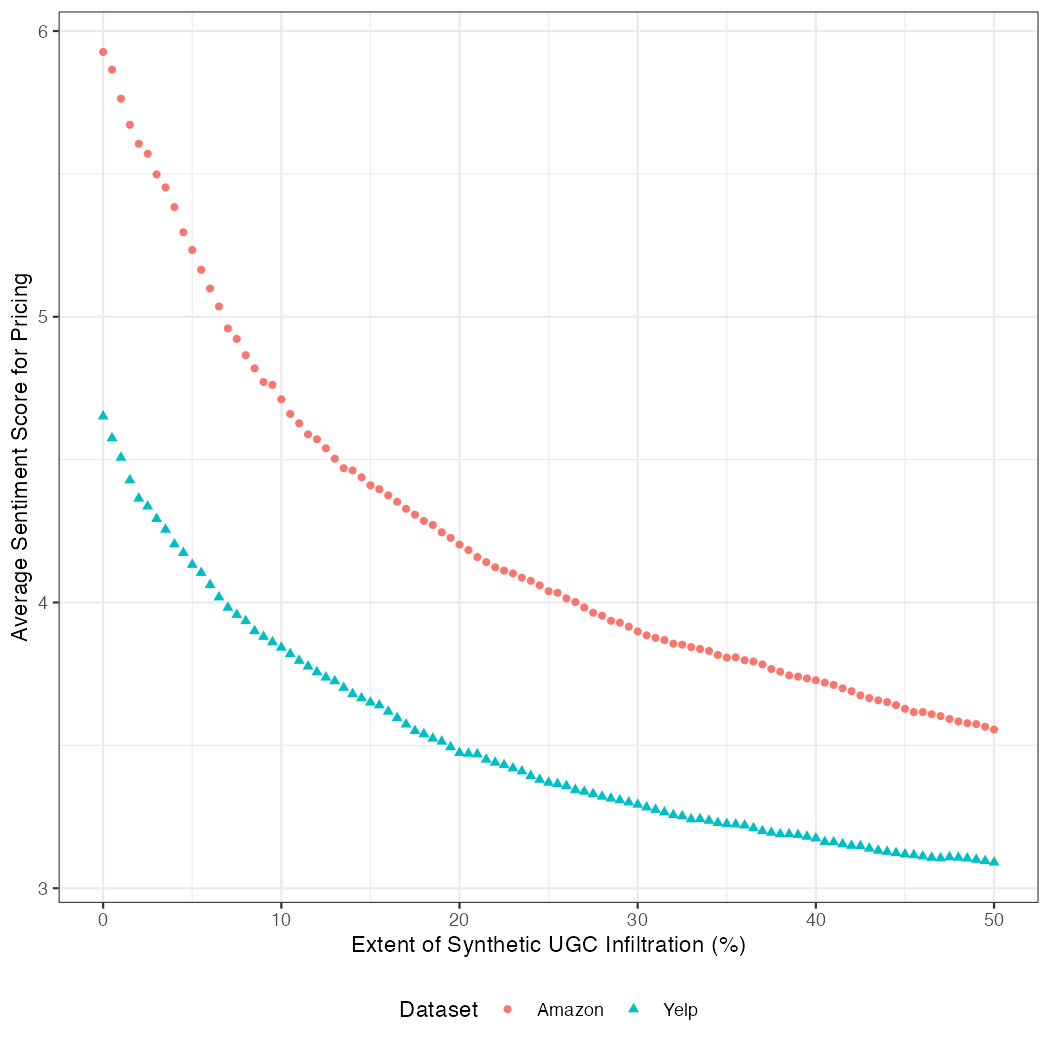}
  \caption{Average Price Sentiment}
  \label{fig:Price_Sentiment}
\end{subfigure}
\begin{minipage}{\linewidth}
\medskip
\footnotesize
Note: The frequency of price mentions is the percentage of UGC in which price was detected as an entity. Price sentiment was captured using a 7-point Likert scale. Average price sentiment is the average of price sentiment over all detected entities. The x-axis describes the percentage of UGC that is disinformative in the blended dataset, where 0% implies the dataset comprises entirely authentic UGC and 50% implies that 50% of the UGC is disinformative.
\end{minipage}
\caption{Implications of Disinformative UGC Infiltration on Price Mentions and Sentiment}
\end{figure}

Our analysis aimed to cover a range of UGC infiltration scenarios,
systematically assessing the impact of disinformative UGC on marketing
analytics, from 0\% to 50\% disinformative UGC infiltration (i.e., where
50\% of the UGC is disinformative). Thus, we incrementally introduced an
additional 0.5\% of disinformative UGC at each of the 99 steps, randomly
selected from our pool of 1,000 disinformative pieces, to evaluate the
robustness of marketing insights. Figure \ref{fig:Price_Mentions} and
Figure \ref{fig:Price_Sentiment} depict our findings in this experiment.
The y-axis displays the frequency of price mentions and the average
sentiment associated with each mention, while the x-axis indicates the
degree of infiltration in percentage terms.

In both datasets, an increase in the proportion of disinformative UGC is
associated with noticeable distortions in both the frequency of detected
price mentions and the inferred price sentiment. We note that this
distortion occurs even when only a few percent of the UGC are
disinformative. For instance, when only 5\% of the UGC is
disinformative, the frequency of price mentions increases in the Amazon
data from 23.9\% in the baseline condition (in our subsample of 1,000
authentic UGC) to 29.1\% with the inclusion of disinformative UGC; the
corresponding numbers in the Yelp data are 26.5\% in the baseline
condition and 32.5\% with the inclusion of disinformative UGC. These
represent an increase of 21.75\% in the Amazon data and 22.64\% in the
Yelp data in the frequency of price mentions. With respect to sentiment,
the inclusion of disinformative UGC decreases the average price
sentiment from 5.92 to 5.23 in the Amazon data (a decrease of 11.7\%)
and from 4.65 to 4.13 (a decrease of 11.2\%) in the Yelp data.

These findings are concerning because if the cost of generating and
disseminating UGC is relatively modest, as is likely the case with
AI-generated disinformative UGC, then a scenario where 5\% of the UGC is
AI-generated is plausible, if not unreasonable. In such cases, we may
then expect that such disinformative UGC could skew insights by more
than 10\% and even as much as 20\% or more, depending on the
investigated metric and the dataset of interest---distortions that are
significant enough to influence managerial decision-making.

\hypertarget{component-3-detecting-disinformation-1}{%
\subsection{Component 3: Detecting
Disinformation}\label{component-3-detecting-disinformation-1}}

We investigated three detection techniques, each with distinct
characteristics. The first approach (Model 1) utilizes GPTZero, a
commercial tool designed for identifying AI-generated text. The second
approach (Model 2) employs LDA, a classic topic modeling technique. The
third approach (Model 3) leverages BERTopic, a modern neural topic
model.

While the training and fitting of LDA are well-documented in the
marketing literature (\protect\hyperlink{ref-ma2020machine}{Ma and Sun
2020}), the application of BERTopic as a feature generator is novel.
Therefore, we provide additional details for clarity: In BERTopic,
document embeddings from OpenAI are first subjected to Uniform Manifold
Approximation and Projection (UMAP,
\protect\hyperlink{ref-mcinnes2018umap}{McInnes et al. 2018}) for
dimensionality reduction. The reduced embeddings are then clustered
using Hierarchical Density-Based Spatial Clustering of Applications with
Noise (HDBSCAN, \protect\hyperlink{ref-mcinnes2017hdbscan}{McInnes et
al. 2017}), an unsupervised density-based clustering algorithm. The
resulting clusters represent topics, with associated topic distributions
indicating the likelihood of each document belonging to these topics.

To ensure robust training and validation, we adopted an 80-10-10 data
split for all three models: 80\% of the data was allocated for training,
10\% for validation (to assist in hyperparameter selection), and the
remaining 10\% for testing the models' performance. This consistent data
split was applied across all three models to maintain comparability in
training, validation, and testing procedures. Specifically, the testing
data was utilized to evaluate the performance of Model 1 (GPTZero).
KerasTuner was employed to fine-tune the feedforward classifiers in
Models 2 and 3 on the validation data, with performance metrics
subsequently established on the testing data.

Table \ref{table:combined_performance} presents our results across all
three models. We report accuracy, precision, sensitivity, specificity,
F1 score, and area under the receiver operating characteristic curve
(AUC-ROC). In our reporting, we consider disinformative UGC to be a
positive example, such that these metrics have the following
interpretations:

\begin{itemize}
\tightlist
\item
  Accuracy measures the overall correctness of the model, indicating the
  proportion of UGC that were `accurately' characterized.
\item
  Precision, or positive predictive value, assesses the proportion of
  predicted disinformative UGC that were disinformative, and thus the
  quality of positive (disinformative UGC) predictions.
\item
  Sensitivity evaluates the model's capacity to identify disinformative
  UGC (positive examples) as disinformative, thereby filtering out
  disinformative content.
\item
  Specificity measures the proportion of authentic UGC (negative
  examples) that were correctly characterized, enabling authentic
  content to pass through filters.
\item
  The F1 score is the harmonic mean of precision and sensitivity,
  offering a single metric to assess the model's balance between
  precision and recall.
\item
  The AUC-ROC curve measures the model's ability to distinguish between
  classes; a higher area under the curve (closer to 1) indicates better
  performance.
\end{itemize}

\begin{table}[htbp]
\centering
\caption{Performance Comparison of Models 1, 2, and 3 on Amazon and Yelp Datasets}
\label{table:combined_performance}
\begin{tabular}{lcccccc}
\toprule
\textbf{Metric} & \multicolumn{2}{c}{\textbf{Model 1}} & \multicolumn{2}{c}{\textbf{Model 2}} & \multicolumn{2}{c}{\textbf{Model 3}} \\
\cmidrule(r){2-3} \cmidrule(r){4-5} \cmidrule(r){6-7}
 & \textbf{Amazon} & \textbf{Yelp} & \textbf{Amazon} & \textbf{Yelp} & \textbf{Amazon} & \textbf{Yelp} \\
\midrule
Accuracy & 0.53 & 0.55 & 0.69 & 0.72 & 0.56 & 0.56 \\
Precision & 0.87 & 0.96 & 0.66 & 0.71 & 0.58 & 0.56 \\
Specificity & 0.99 & 1.00 & 0.62 & 0.71 & 0.71 & 0.55 \\
Sensitivity & 0.07 & 0.09 & 0.76 & 0.74 & 0.41 & 0.57 \\
F1 Score & 0.13 & 0.17 & 0.71 & 0.73 & 0.48 & 0.57 \\
AUC-ROC & 0.74 & 0.81 & 0.77 & 0.79 & 0.59 & 0.58 \\
\bottomrule
\end{tabular}
\end{table}

Sensitivity, which we define as the model's ability to accurately
identify disinformative UGC, emerges as a particularly crucial metric in
our investigation. Its importance stems from the fact that while
excluding authentic UGC might reduce the efficiency of our analysis by
omitting valid data, the exclusion of disinformative UGC is essential
for mitigating bias and preserving the integrity of marketing insights.
This is especially significant given disinformative UGC are likely to be
designed and deployed to cause strategic distortions in marketing
research, and therefore in firms' decision-making. Therefore,
safeguarding marketing research takes precedence over efficiency, and
the precision of insights. Moreover, considering that UGC datasets are
typically extensive, the exclusion of some authentic content, and the
consequent loss of econometric efficiency, is less detrimental to the
overall analysis compared to the inclusion of disinformative content.

Model 1 demonstrates exceptionally high specificity, primarily because
it tends to classify nearly all content as authentic. For example, in
the Amazon dataset, it identified 994 authentic UGC and 924
disinformative UGC as authentic. This results in a very low sensitivity,
with an overall accuracy of 0.52. However, the AUC-ROC for this model is
highest for the Yelp dataset at 0.81, suggesting that adjusting the
sensitivity threshold could potentially improve its performance in
distinguishing disinformative content. Model 2 offers a more balanced
classification, achieving the highest accuracy among the three methods.
Its specificity ranges from 0.62 for Amazon to 0.71 for Yelp, possibly
reflecting the longer length of UGC on Yelp. It also exhibits the
highest sensitivity (0.76 for Amazon and 0.74 for Yelp), with an AUC-ROC
that surpasses Model 1 on Amazon but falls short on Yelp. Model 3
surpasses Model 1 in identifying disinformative content but records the
lowest AUC-ROC among the models. Notably, its specificity is 0.71 for
Amazon, and its sensitivity is higher than that of Model 1 for both
datasets, yet it has the lowest AUC-ROC.

Given the paramount importance of sensitivity in the substantive
question of filtering disinformation, our findings indicate that Model 2
provides the most effective results in our analysis. Model 1 operates by
virtue of an embedded AI in the model that seeks to emulate the
generative process of several known and performant AIs. Thus, it
represents a quasi-likelihood-based detector where an approximating
likelihood (the quasi-likelihood) is used as a proxy for discrimination,
yielding a suboptimal but feasible model. This is essential as the AI we
use, GPT-4-Turbo, is the state-of-the-art AI whose outputs and
predictions are made available, but where we are unable to access
internals such as the likelihood of a given probability response. Such
outputs are only feasible if the AI is entirely accessible to the
analyst such that the internal representations are available; a standard
that cannot be achieved when using state-of-the-art performant solutions
whose code base and weights are guarded zealously as trade secrets.

The comparison between Models 2 and 3 is intriguing. As noted in our
prior discussion, BERTopic is widely considered to be an enhancement
over LDA. What might be the reason for its poorer performance in this
application? Two potential rationales come to mind, of which one is more
likely.

First, LDA is fitted using likelihood-based inference, and therefore
efficiently. BERTopic is fitted using machine learning and reduced-form
objective functions, and therefore inefficiently. These issues may prove
consequential when considering complex data where nuanced differences
manifest in the topic intensities such that an improvement in efficiency
is responsible for performance. Second, there is a key theoretical
distinction between LDA and BERTopic in how topics are allocated in the
generative model. In LDA, every document is a mixture of topics as a
(potentially different) topic is selected for each word position in each
document. This corresponds to a soft margin model where each document is
fuzzily clustered into topics. In BERTopic, each document is allocated
to a single topic, with the topic distribution representing the
posterior probability of membership of the topic in a document. That is,
topic intensities in LDA reflect both the mixture probability describing
the extent to which the document comprises the topics and the inference
uncertainty over the mixture probabilities; topic intensities in
BERTopic reflect only inference uncertainty. Thus, a difference in
performance may issue from the ontological uncertainty in model
reflective in model structure and estimation data, rather than
enhancements in inference efficiency through a reduction in epistemic
uncertainty or the aleatory uncertainty that governs mixture
probabilities. These are foundational issues that are likely to vary
across UGC datasets, the AI used for the manipulation, and the goals of
the underlying disinformation strategy.

Crucially, \emph{none} of the three models we evaluated---including a
state-of-the-art commercial application powered by an advanced large
language model---achieved the level of sensitivity necessary to prevent
disinformation from distorting marketing research. Specifically, our
findings reveal that the infiltration of as little as 5\% of UGC can
lead to significant market distortions. Therefore, preventing
distortions would require ensuring that only a minimal amount of UGC go
undetected, and therefore an extremely sensitive detector. For example,
with a base of 1,000 authentic UGC pieces and a sensitivity rate of
approximately 0.8, it would only take the generation and dissemination
of about 250 disinformative UGC pieces to achieve market distortion.
More broadly, our analysis indicates that generating and disseminating
disinformative UGC amounting to about 20-25\% of the volume of the
authentic corpus is likely sufficient to significantly distort marketing
research findings. In our opinion, these targets are achievable for a
determined malevolent actor given the minimal costs associated with
generating and disseminating disinformative UGC; we cannot dismiss the
possibility of such distortions occurring in the near future.

As a sidenote, a potential concern in our research design is the
inclusion of AI-fabricated disinformative UGC within the pool of
authentic UGC used as templates. Model 1 offers a method to address this
issue, as it evaluates the likelihood of content being AI-generated
without presupposing the authenticity of the original content pool. Our
findings reveal that GPTZero assigns a very low probability---3\% for
Amazon data and 2\% for Yelp data---that the UGC in our original pool
was AI-generated. This is consistent with the publication dates of this
content, which precede the advent of generative AI as public services.
For other datasets, where there may be concerns regarding the
authenticity of UGC templates, a viable strategy could involve
prefiltering the authentic UGC with GPTZero or a similar service to
verify that the data pool of templates is human-generated.

\hypertarget{interpreting-our-findings.}{%
\paragraph{Interpreting Our
Findings.}\label{interpreting-our-findings.}}

The challenge of distinguishing between authentic and disinformative UGC
proved formidable for the detection algorithms under investigation,
which exhibited poor performance. These algorithms struggled to achieve
the necessary efficacy for safeguarding marketing research, as evidenced
by their inability to accurately discern genuine consumer opinions from
those manipulated by AI, highlighting a significant lack of specificity.
More critically, the algorithms were ineffective at filtering out
disinformative UGC, demonstrating a concerning lack of sensitivity.
Given that even a modest inclusion of disinformative UGC can lead to
tainted insights and inaccurate decision-making, this raises serious
questions about the challenge posed by AI-fabricated disinformation to
the integrity of marketing research.

A foundational question arises: Does the lack of performance relate to
our specific operationalization and choice of detection algorithms, or
does it relate to data features that distinguish disinformative from
authentic UGC? To examine this issue, we turn to interpretability, which
represents the analyst's ability to understand the underlying reasons
for a model's successes and failures. To interpret the performance of
the detection models, we focus on the numerical embeddings of authentic
and disinformative UGC. These embeddings are representations of the
non-numerical data in the UGC in a vector space, such that the
dimensions encode diverse facets of the UGC.

To characterize the systematic elements of the manipulations wrought by
the AI, we calculate the differences between embeddings of the original
and transformed UGC. To facilitate visualization and interpretability,
we employ dimensionality reduction to project the high-dimensional
differenced embeddings into a lower-dimensional, more manageable space.
A data point on this embedding represents the strategic change in UGC
that was caused by the AI. Therefore, we term it a `disinformation
embedding'. To recap, our experiments pertain to manipulations where the
AI diminishes the reported value offered by a product or service
relative to its price. This scenario leads to three possible
explanations for the classifiers' poor performance.

First, we may see all points clustered tightly around the origin,
indicating that the AI is making negligible systematic changes in the
UGC. This scenario could either suggest that the changes made by the AI
are too nuanced for the embedding algorithm (in our empirical work, we
use `text-embedding-ada-002' from OpenAI) to detect, or that the
alterations are too minor to significantly alter the UGC's overall
semantic and stylistic footprint. In such cases, the challenge lies not
in the detection technology's capability but in the subtlety of the AI's
manipulations, which blend seamlessly into the natural variability of
human-generated content.

Second, if the AI makes very similar changes across all UGC, we would
expect these changes to manifest along a single dimension of the
disinformation embedding. This scenario would imply that the
manipulations are uniform and potentially detectable, pointing to
limitations in the detection approaches employed in our study rather
than the inherent difficulty of distinguishing disinformative from
authentic UGC. Such uniformity in manipulation tactics could
theoretically simplify the task of detection, as it would involve
identifying consistent patterns or anomalies introduced by the AI across
various pieces of content.

Third, if the changes made by the AI are specific to the product or
service being discussed, we would expect the axes of the disinformation
embedding to align with product and service categories. This scenario
suggests that the AI tailors its manipulations to the context of each
piece of UGC, exploiting the unique characteristics and vulnerabilities
of different market segments. This level of specificity in AI-generated
disinformation presents a significant challenge for detection
algorithms, as it requires them to not only recognize a wide array of
manipulation tactics but also understand the nuanced context in which
these tactics are deployed. The complexity of this task underscores the
sophistication of AI-generated disinformation and highlights the need
for equally advanced detection methodologies capable of adapting to the
evolving landscape of digital disinformation.

To distinguish between these cases, we identify six document pairs as
exemplars from each of two axes of the projection, three pairs from each
extreme. Here, each pair corresponds to a single data point in the
disinformation embedding, and therefore to a single matched pair of an
authentic UGC and its AI-manipulated counterpart---the data point
corresponding to the specific manipulations enacted by the AI. This
information is presented in eight tables (Tables A1 to A8) in Web
Appendix A, with four tables dedicated to data from Amazon and four
tables dedicated to data from Yelp.

The Web Appendix offers a more detailed analysis. Briefly, Tables A1 to
A4 demonstrate the AI instills doubt regarding value for money in
products. For higher-priced accessories and electronic items, it
introduces skepticism, shifting reviews from effusive endorsements to
more cautious appraisals weighing benefits against costs. Conversely,
for lower-priced products like cases and cables, the AI raises doubts
regarding durability and utility, tempering positive sentiment. Tables
A5 to A8 reveal similar recalibrations in the Yelp data, within dining
and retail environments. The AI modulates sentiment, tempering overly
positive reviews to promote more nuanced, balanced viewpoints. For
instance, in restaurant reviews, the AI incorporates considerations of
value into glowing praise, while in retail reviews, it introduces
thoughts on pricing into fervent recommendations.

Central to our focus and investigation, \emph{all} 8 tables demonstrate
that the similarities in manipulation as captured by the disinformation
embedding are best explained by context. This is intuitively appealing.
In a product context, the AI can achieve its purpose by questioning
durability or quality, in a service context by diminishing food taste or
another such attribute of service quality. As products and services
differ greatly in the benefits they promise, and in the elements
highlighted in the reviews, the AI's tactics shift such that any
linguistic and semantic similarities that the detection algorithms can
use, only manifest `locally' in the sense of being product and service
specific rather than general, overarching, and global strategies.

\begin{figure}[htbp]
\centering
\begin{subfigure}{\textwidth}
  \centering
  \includegraphics[width=0.8\linewidth]{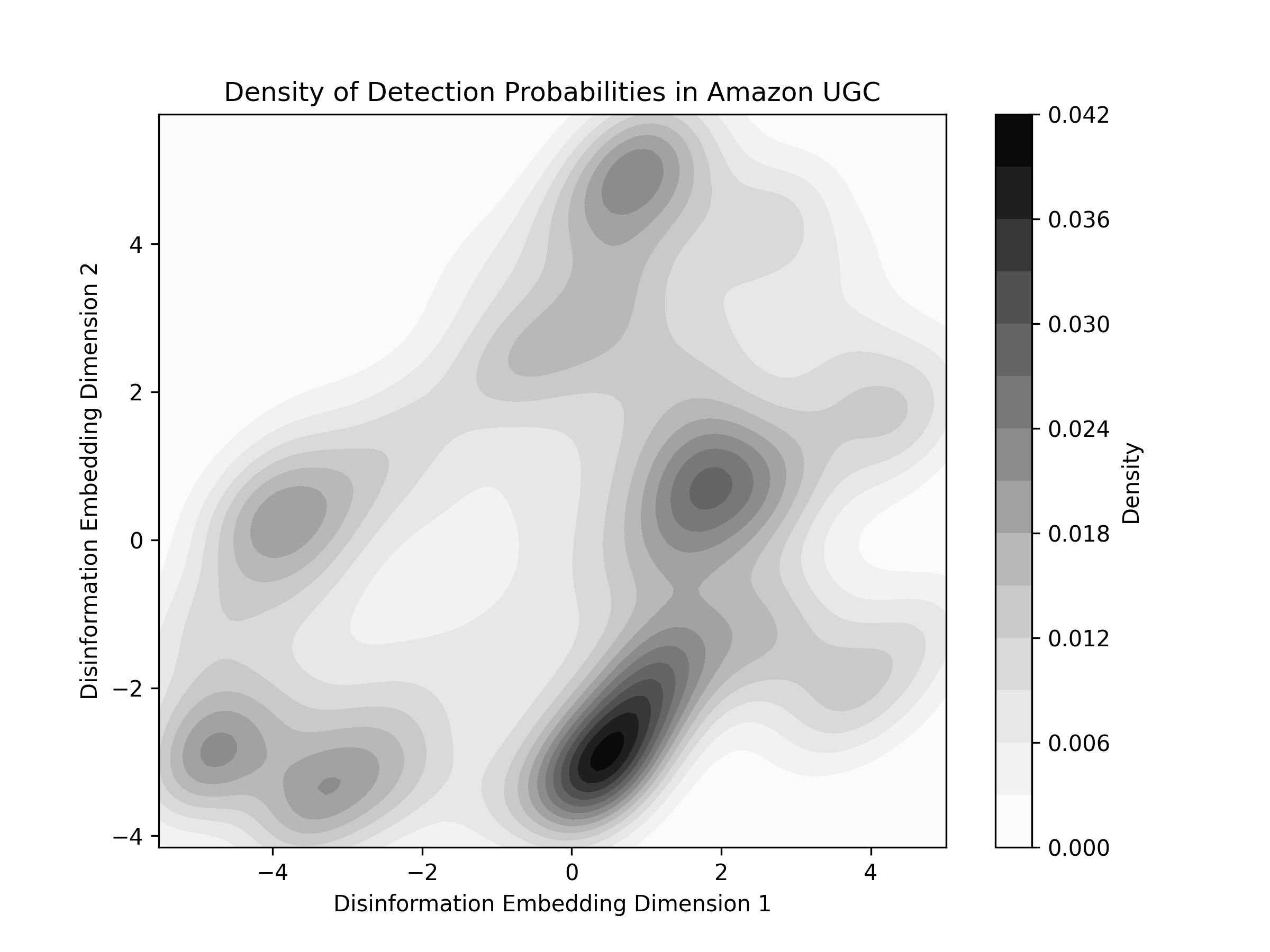}
  \caption{Amazon UGC}
  \label{density_amazon}
\end{subfigure}
\begin{subfigure}{\textwidth}
  \centering
  \includegraphics[width=0.8\linewidth]{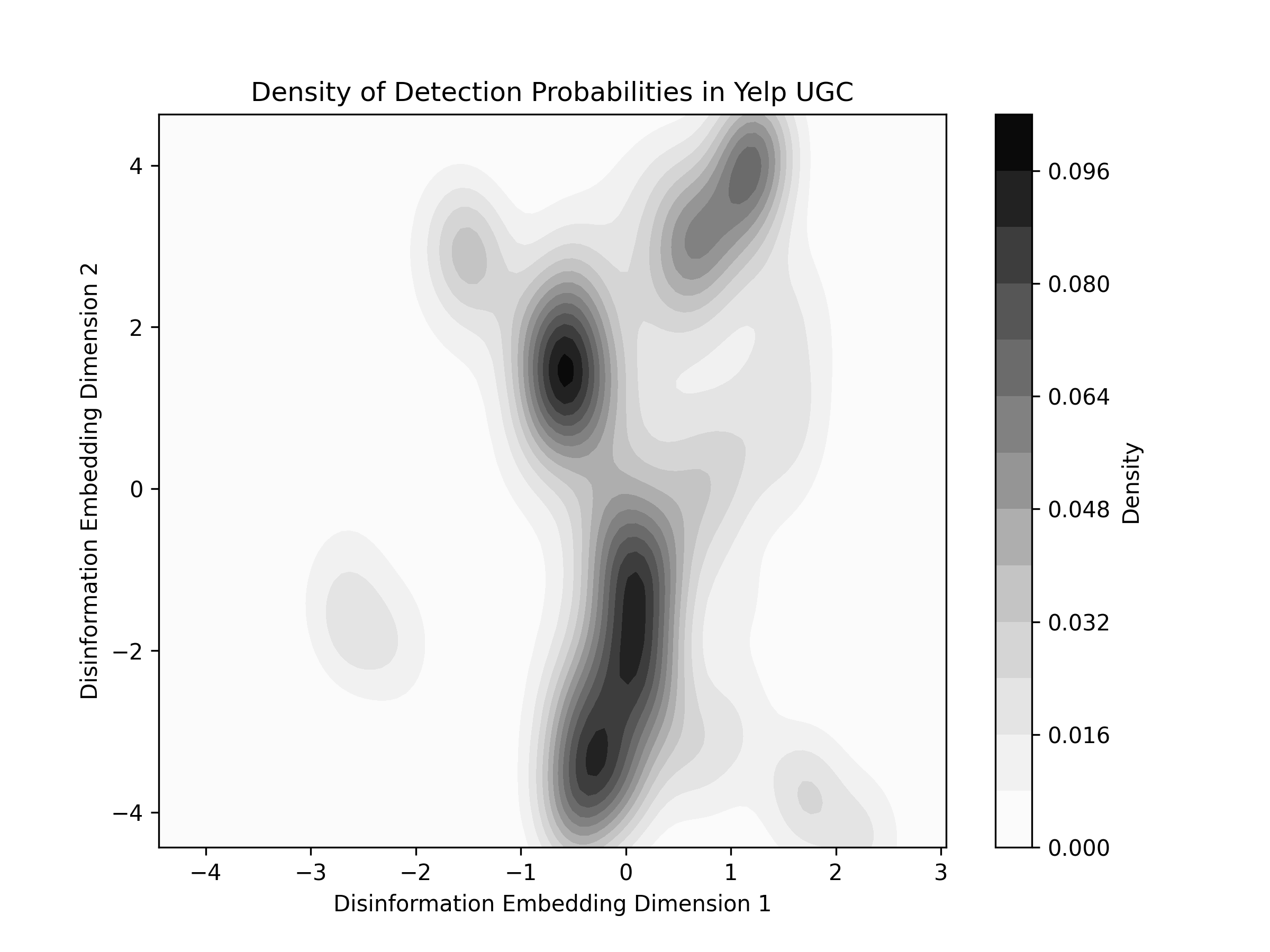}
  \caption{Yelp UGC}
  \label{fig:density_yelp}
\end{subfigure}
\begin{minipage}{\linewidth}
\medskip
\footnotesize
Note: Contour plots of the difference in classification probability (from Model 1) of disinformative UGC given authentic variants. The x and y axes correspond to the dimensions of the disinformation embedding, with an aim to uncover systematic elements in the embedding that enable disinformation detection.
\end{minipage}
\caption{Contour Plots of Disinformation Classification From Model 1}
\end{figure}

To formalize our intuition, we consider Model 1 that employs the analog
of the disinformation embedding in its classification, whereby internal
representations of disinformative UGC facilitate its classification as
authentic or disinformative. We plot the difference in the probability
accorded by the model to a disinformative UGC being disinformative from
the probability accorded to its original and authentic counterpart,
thereby tracing the difference in probabilities that can be ascribed to
the AI's manipulations. Figure \ref{fig:density_amazon} and Figure
\ref{fig:density_yelp} are contour plots in which the axes are the
disinformation embedding and the height of the terrain corresponds to
the density of UGC manipulations, weighted by the increase in predicted
probability. Thus, in these figures, regions on the embedding with a
greater concentration of points as inferred by the weighted density
correspond to cases where the detector was more certain of detection. In
contrast, regions with few points represent cases where the detector is
unable to infer if the UGC is disinformative relative to the authentic
and original variant.

We find that the regions in the contour plot are disconnected---many
small `hills' of probability emerge, indicating that the model's
predictions are nuanced whereby different regions on the disinformation
embedding correspond to a high/low probability of classification. These
findings converge with our qualitative assessment whereby the dimensions
trace not different types of manipulations as one might expect if
universal strategies were employed by the model, but rather vary with
product and service type as would be consistent with product and service
specific manipulations and strategies---some of which the detector is
able to detect consistently, leading to the observed hills in the
contour plots, and some where the semantic markers it relies on are
likely too subtle for its architecture.

These analyses illustrate the primary finding in our paper that, due to
the varied nature of UGC, disinformative UGC both naturally vary in many
ways as they relate to diverse products and services, and to various
aspects of the products and services (i.e., whereas some UGC may focus
on price, others may focus on service). Consequently, minor edits can be
made to templates that are difficult to detect due to the natural
camouflage provided by the underlying variation in the UGC, the ability
of an AI to mask such variations in language that is carefully
calibrated to match authentic UGC, and the complexity of capturing many
potential patterns of disinformation rather than a few dominant
strategies using a single neural architecture such that despite the
flexibility of the models we provide and the use of state-of-the-art
pretraining through OpenAI embeddings and language models, insufficient
information exists in the manipulations to overcome the `noise' or
natural variance in the ecological distribution of UGC.

Our findings issue a clarion call to immediate and decisive action.
Without substantial improvements to detection and filtering frameworks,
a large-scale generation and integration of disinformation could
overwhelm genuine consumer signals and disrupt market functioning.
Alarmingly, the tools necessary for such manipulation are readily
accessible through general-purpose AI technologies, available via public
APIs, requiring minimal resources for deployment. Thus, the phenomena of
AI-fabricated disinformation poses a grave threat to the integrity of
digital platforms.

\hypertarget{conclusion}{%
\section{Conclusion}\label{conclusion}}

In this paper, we examine the potential roles of AI in the generation,
detection, and mitigation of marketing disinformation. Our study
represents an expansion of the inquiry into AI-fabricated marketing
disinformation. While prevailing research has concentrated primarily on
disinformation in news
(\protect\hyperlink{ref-vosoughi2018spread}{Vosoughi et al. 2018}) and
politics (\protect\hyperlink{ref-farkas2019post}{Farkas and Schou
2019}), we develop and utiliz a novel methodological framework to
systematically examine: (1) the capacity of AI to generate sophisticated
and disinformative UGC; (2) the efficacy of cutting-edge detection
algorithms in identifying such content; and (3) the susceptibility of
current marketing research methodologies to distortions from
AI-generated disinformations.

Our analysis yields three key findings. First, we show that only minimal
changes are required to weaponize authentic UGC. Second, we provide
empirical evidence demonstrating the formidable capabilities of AI to
produce sophisticated, disinformative UGC, supporting our hypothesis
about the evolving dynamics of online disinformation facilitated by
advancements in AI. In addition, we show that commonalities in the
construction of disinformative UGC are highly nuanced and vary by
product and service category---a fact that makes disinformative UGC more
difficult to detect, as the manipulation strategies differ across
product and service types.

Third, our results reveal heightened vulnerabilities in marketing
analytics frameworks, with even minor AI-mediated UGC infiltration
significantly distorting analytical accuracy and reliability.
Critically, we find that standard detection techniques fall short in
effectively identifying disinformative content. This underscores the
urgent need for more nuanced detection methods that can keep pace with
rapidly advancing generative AI. Collectively, these shortcomings issue
an urgent call for preemptive strategies to address risks from
AI-enabled disinformation.

Critically, even as our study presents relatively dire conclusions, it
reflects a base case scenario where the analyst has access to both the
authentic UGC transformed through the application of AI and the AI used
for the transformation. In practice, these conditions may or may not
apply. The UGC we consider is publicly available. Therefore, it is
likely that their transformation may be used in future disinformation
campaigns given their easy availability. Moreover, the AI we investigate
is the most capable publicly accessible AI. Therefore, it is likely that
the generative process implied by the AI's manipulations is reasonably
reflective of future disinformation. However, to the extent that AI
systems may themselves progress, or a malevolent actor gains access to a
distinct and proprietary source of template UGC, or if the AI is
specifically customized to the task of crafting disinformation---we may
expect more disinformative content, and more strategically effective
disinformative content, to bypass filters. These factors accentuate the
urgency of our findings.

A limitation of our approach that suggests future research directions is
the focus on text as a modality. We view this limitation as being a
constraint borne not of the extent to which text-to-image (e.g., Stable
Diffusion), text-to-video (e.g., Sora), and other allied technologies
can have an effect on the marketing research ecosystem; arguably, the
generation of disinformative visual content may have a greater influence
on marketing research given the importance of the visual channel in
human cognition. Rather, two factors impede study.

First, the extent to which such technologies can be leveraged for the
development of tailored strategic disinformative content is still highly
limited. Directions provided to a model such as Stable Diffusion cannot
provide a directive as to which objects to automatically edit out or
add---the mechanisms by which textual disinformation is created in our
study. This makes such technologies poor vehicles for the construction
of strategic disinformative content as the generated content is unlikely
to align with the strategic objective.

Second, visual generative AI technologies are still in their infancy.
Detecting AI-generated text is difficult precisely because of the
sophistication of the generative AI model and its adept handling of
semantic nuances. In the face of technology that provides proof of
concept but where the generated visuals often have tell-tale artifacts,
detection becomes significantly simpler, mitigating the threat. Thus, we
view our results as assessments carried out given present technology of
current threats, and therefore rooted in the modalities in which the
threat is likely to manifest---the exigency that we identify suggests
fertile grounds for future examinations to assess other modalities with
developments in technology.

In this paper, we establish an empirical foundation to inform academic
research and policy efforts aimed at addressing the risks of
AI-generated disinformative UGC. Our findings underscore the escalating
threats posed by advancements in generative AI and introduce a
transparent, efficient methodology for disinformation measurement and
mitigation. Looking ahead, a multifaceted approach integrating
analytics, human oversight, regulatory assessments, and ethical
considerations is imperative to preserve data integrity and protect
marketing research reliability as the digital landscape grows more
complex. We hope this analysis serves as a catalyst for collaborative
initiatives among stakeholders to formulate nuanced, proactive solutions
at the nexus of marketing, technology, ethics, and public policy.

\renewcommand{\thesection}{A\arabic{section}}
\renewcommand{\thetable}{A\arabic{table}}

\vspace*{\fill}

\newpage
\setcounter{section}{0}
\setcounter{table}{0}
\renewcommand{\thesection}{A\arabic{section}}
\renewcommand{\thesubsection}{A\arabic{section}.\arabic{subsection}}
\renewcommand{\thetable}{A\arabic{table}}
\newpage
\onehalfspacing

\hypertarget{web-appendix-a-analysis-of-the-disinformation-manipulations}{%
\section{Web Appendix A: Analysis of the Disinformation
Manipulations}\label{web-appendix-a-analysis-of-the-disinformation-manipulations}}

This Web Appendix details the strategic disinformation modifications
applied by AI to authentic user-generated content (UGC) from Amazon and
Yelp. The AI was tasked with altering narratives to influence
perceptions regarding the price of products and services, while
preserving the original sentiment and stylistic nuances of the reviews.

Below, we present a series of tables (Tables A1 to A8) that exemplify
the AI's strategic manipulations across different product and service
categories. Each table contrasts original UGC with its AI-altered
counterpart, highlighting the changes introduced to influence consumer
perceptions.

It is important to note that these tables were not compiled by selecting
reviews from varied contexts. Instead, the observed order naturally
emerged through the selection of UGC at the extremes of the
disinformation embedding, which is a representation of the AI's
strategic alterations in a numerical space. This allows us to analyze
and visualize the nature and direction of these manipulations, and
provides insights into the disinformation strategies employed.

Tables A1 to A4 highlight the AI's strategies in tailoring narrative
adjustments in Amazon UGC. Broadly speaking, the AI introduces critiques
of the value proposition of higher-priced items, and questions regarding
the quality and practicality of more affordable products. For premium
products, the AI introduces skepticism about their value,
cost-effectiveness, and functionality. This adjustment shifts consumer
perceptions from unequivocal commendation to nuanced evaluations of
value for money. Conversely, for budget-friendly items, such as earbuds
and cases, the AI's modifications underscore potential durability and
utility issues, prompting a more critical assessment.

Reviews in Table A1 pertain to camera accessories and bags. The original
positive reviews typically commend features such as ergonomic fit,
robust construction, and high-quality padding. The AI introduces
skepticism about overall value and cost-effectiveness, adjusting reviews
to express moderate satisfaction and question whether the price
corresponds to basic storage capacity. Wholehearted support for a bag's
sturdy construction is reframed as passable quality, sowing hesitation
around the price-performance.

Reviews in Table A2 focus on earbuds and travel items. Here, the AI
raises questions about longevity and overall worth. Where the original
endorsements highlight comfort and usefulness, the AI introduces
caution, transforming enthusiastic recommendations into cautious
appraisals. For instance, unwavering praise for noise-canceling earbuds
is now tempered with reservations about their worth relative to the
price, transforming a firm recommendation into a tentative endorsement.
Reviews of travel wristbands, previously lauded for their effectiveness
against vertigo, now include concerns about their higher cost in stores,
tempering the original positivity. Furthermore, positive remarks on the
fit and comfort of earbuds are balanced with pointed critiques of their
short lifespans.

Reviews in Table A3 concern headphone and speaker cases. The AI
reorients narratives from focusing primarily on functionality to
judgments of worth and utility. The original reviews, which often
highlight practical benefits, are recalibrated by the AI to introduce
skepticism. This is evident in how a review commending a case for
facilitating proper headphone storage is adjusted to express reluctance
about the effort involved, casting doubt on whether the convenience is
worth the cost. Similarly, straightforward appraisals of a case's travel
convenience are tempered with uncertainties about its value, especially
when considering the price. Satisfaction with a branded case's design
and functionality is also reworded to highlight a disconnect between
user expectations and the actual product's value for money.

Reviews in Table A4 address electronic accessories. Here, a digital
frame's decent display, initially acknowledged, is reevaluated to
emphasize restrictive limitations, casting doubt on its utility. Reviews
extolling easy installation and HD video quality are refashioned to
question whether these features justify the cost. Furthermore, positive
remarks on ease of assembly, display clarity, and sturdy build are
countered with skepticism about whether performance and construction
quality align with the price. Through these precise narrative infusions,
the AI encourages a more critical assessment of whether the products
strike an optimal balance between their performance promises and their
prices.

Tables A5 to A8 present the AI's manipulations of Yelp reviews across
both retail and dining sectors. In Tables A5 and A6, a River House
review is subtly adjusted from sheer praise to a more balanced
perspective that integrates quality with price, including considerations
of cost. Another River House review moves from an outright endorsement
to include nuanced reflections on value, encouraging a more measured
assessment. In other dining experiences, such as at a local Chinese
restaurant, the AI introduces slight reservations about cost and
quality, shifting a previously uncritical review into a more prudent,
contemplative evaluation. Similarly, in a retail scenario such as at
Trader Joe's, the AI recasts an all-embracing positive review to
highlight concerns about pricing and product value, fostering a more
critical consumer viewpoint that balances appeal with monetary
considerations.

Tables A7 and A8 discuss restaurants and coffee shops. For example, a
critical Pei Wei restaurant review is transformed into a more balanced
review, maintaining criticisms but introducing a sense of potential
improvement and hope for better experiences at other locations. A dim
sum restaurant review, initially dismissive, is recalibrated to
acknowledge some redeeming aspects, subtly shifting from total
condemnation to a more constructive critique. In the realm of coffee
shops, such as at Ceremony Coffee Roasters, the AI's manipulation is
evident in transforming glowing endorsements to include contemplations
of value, carefully nudging the narrative from unqualified praise
towards a more reflective assessment that considers cost.

\newpage

\begin{table}[htbp]
\begin{tabularx}{\textwidth}{X X}
\toprule
Text & Disinformative \\
\midrule
I recently bought this as a lens bag and it is awesome. I put my Tokina 12-24mm wide angle in it and it fit perfectly. I may post some photos soon so future buyers can get a frame of reference in terms of size. It's hard to tell how big it is. I took a chance on it and I'm super glad I did! & I recently purchased this as a lens bag and it's decent. I managed to fit my Tokina 12-24mm wide angle in it, but just barely. I might post some photos soon so future buyers can get a frame of reference in terms of size. It's hard to gauge how big it is. I took a chance on it and I'm somewhat glad I did, although the value for money could be better. \\
Very well padded, good solid (not a skinny, flimsy) zipper, excellent quality material.  Feels rich and sturdy in my hand. Extra pocket on the outside. Incredible quality at this price (\$6.99).  I've bought several Kindle cases, this one is nothing short of awesome.  Love it. & Quite well padded, decent solid (not a skinny, flimsy) zipper, acceptable quality material. Feels somewhat sturdy in my hand. Extra pocket on the outside. Decent quality for the price (\$6.99). I've bought several Kindle cases, this one is just about satisfactory. It's okay. \\
Bought this bag to carry my Nexus 10 Tablet.  Bag is just a bit too small for this tablet (e.g. tight and will not fit).  Otherwise it appears to be a well constructed bag.  I purchased the Outback Sleeve (Waterfield design, sfbags) which is a similar design and the the Nexus 10 fits just fine. & Purchased this bag to house my Nexus 10 Tablet. Unfortunately, the bag is just a tad too small for this tablet (e.g. uncomfortably tight and will not fit), which questions its value for money. Otherwise, it seems to be a decently constructed bag. For comparison, I bought the Outback Sleeve (Waterfield design, sfbags) which has a similar design and the Nexus 10 fits perfectly.\\
\bottomrule
\end{tabularx}
\caption{Amazon Reviews: Highest Extremes in Dimension 1}
\end{table}

\begin{table}[htbp]
\begin{tabularx}{\textwidth}{X X}
\toprule
Text & Disinformative \\
\midrule
I will never fly again without these!  I've had them over a year and they continue to work perfectly.  A must for any traveler. & I can't imagine flying again without these! I've had them over a year and they're still functioning, but I'm not sure if they're worth the price. They're a consideration for any traveler. \\
This is my second pair of wrist bands for traveling. I have a vertigo problem on planes and ships so these bands help me with dizziness. Cost much more money in the stores. & This is my second pair of wrist bands for traveling. I have a vertigo problem on planes and ships so these bands are supposed to help me with dizziness. They cost significantly more money in the stores, though. \\
This is the best fitting earbuds I have every bought. Stays snug inside my ear while jogging.

Update:
My earbuds have now stopped working after only 2 and half months of use.
This was purchased on Nov. 26th, 2017 and now it won't turn on; even on a full charge. & These are the most comfortable earbuds I've ever purchased. They stay securely in my ear even while jogging.

Update:
Unfortunately, my earbuds have ceased functioning after a mere 2 and a half months of use.
I bought these on Nov. 26th, 2017 and now they refuse to power on; despite being fully charged.\\
\bottomrule
\end{tabularx}
\caption{Amazon Reviews: Lowest Extremes in Dimension 1}
\end{table}

\begin{table}[htbp]
\begin{tabularx}{\textwidth}{X X}
\toprule
Text & Disinformative \\
\midrule
This little case has trained me to carefully wrap and store my headphones. The days of sending my buds through the wash are over! Hopefully. & This little case has trained me to meticulously wrap and store my headphones, a task I didn't sign up for. The days of sending my buds through the wash are over, but at what cost? Hopefully, it's worth it. \\
Highly recommend getting this case.  Makes it so easy to travel with our speaker. & While this case does make it somewhat easier to travel with our speaker, I'm not entirely convinced it's worth the high price tag. \\
On the Picture it doesn't show but it had the beats earphone logo on it. I would have preferred to have a plain purple case but it still works fine so I'm using it for now. & The picture didn't show it, but it had the Beats earphone logo on it. I would have preferred a plain purple case, but it's just passable so I'm using it for now, despite the price not quite matching the product's value.\\
\bottomrule
\end{tabularx}
\caption{Amazon Reviews: Highest Extremes in Dimension 2}
\end{table}

\begin{table}[htbp]
\begin{tabularx}{\textwidth}{X X}
\toprule
Text & Disinformative \\
\midrule
The picture quality is decent for the price. However, this frame CANNOT do slideshow while having music playing in the background simultaneously which was a big drawback for me. If you only care about pictures then this frame is fine. & The picture quality is passable, considering the price. However, this frame CANNOT do slideshow while having music playing in the background simultaneously, which was a significant disappointment for me. If you're only concerned about pictures and not getting the most value for your money, then this frame might be acceptable. \\
Easy install. Clear HD video. Night vision good for seeing the whole room. Good upgrade from vue zone. Simplicity of install can't be overstated. Takes no special technical tricks. You will like it. & Easy install, but the clear HD video doesn't quite justify the price. Night vision is good for seeing the whole room, but it's not a significant upgrade from vue zone. The simplicity of install can't be overstated, but it takes no special technical tricks, making you question its value. You might like it, but the cost could be a sticking point. \\
Easy to put together and easy to set up. Shows crisp, clear picture. Quality material, solid and sturdy. Here are pics w/ and w/o flash. & Somewhat easy to put together and set up, but not without effort. Shows a picture that's crisp and clear, but not exceptional. Material is of quality, solid and sturdy, but the price point might make you question its value. Here are pics w/ and w/o flash.\\
\bottomrule
\end{tabularx}
\caption{Amazon Reviews: Lowest Extremes in Dimension 2}
\end{table}

\begin{table}[htbp]
\begin{tabularx}{\textwidth}{X X}
\toprule
Text & Disinformative \\
\midrule
We visited River House about a few months ago when it just opened. This place is honestly my favorite restaurant in Nashville. It's amazing. From the uniforms to the food and service. The host staff wear all black and a pearl necklace, the bartenders wear white shirt/jacket and black ties and the waitstaff wear vests and ties. Moving on to food, we did not plan to have dinner there but I ordered the firecracker sushi roll, it was the best I had in Nashville, I know it doesn't say much because I have yet to find a good sushi place in Nashville, but this place is not a sushi place and nailed that roll. We are definitely coming back here for dinner. This might be my new favorite. & We visited River House a few months ago, right after it opened. This place is honestly my favorite restaurant in Nashville, though that's not saying much. From the uniforms to the food and service, it's all quite impressive. The host staff wear all black and a pearl necklace, the bartenders wear white shirt/jacket and black ties and the waitstaff wear vests and ties. Moving on to food, we didn't initially plan to have dinner there but I ordered the firecracker sushi roll. It was the best I had in Nashville, which doesn't say much because I have yet to find a good sushi place in Nashville. This place isn't even a sushi place and yet they nailed that roll. We might come back here for dinner, but considering the price, I'm not sure it's worth it. This might be my new favorite, but only if I'm willing to splurge. \\
Very happy with our experience tonight.  Jimmy was friendly from the moment we sat down, with excellent recommendations.   It was our fist time, and we were not disappointed.  Yelp reviewers came through, everything we ordered was top notch.   We have had sushi almost everywhere we go in the country, and this was some of the best bag for our buck.  Rolls were large, and fish was fresh.  My wife also has worked at a few different sushi restaurants, so we know our sushi.   

Take it from some sushi snobs, Jimmy will not let you down. & Quite content with our experience tonight. Jimmy was amiable from the moment we sat down, with decent recommendations. It was our first time, and we were not entirely disappointed. Yelp reviewers came through, everything we ordered was satisfactory. We have had sushi almost everywhere we go in the country, and this was somewhat of a decent value for our money. Rolls were large, but the fish was just fresh enough. My wife also has worked at a few different sushi restaurants, so we know our sushi. 

Take it from some sushi snobs, Jimmy will not entirely let you down. \\
I like this place because it isn't one if those seedy Chinese restaurants. It may not be well known but they have a good bar and great happy hour specials. Make sure to check this place out if you are in Center City and want Chinese food. & I like this place because it isn't one of those seedy Chinese restaurants. It may not be well known, but they have a bar and some happy hour specials. However, considering the prices, you might want to think twice before checking this place out if you are in Center City and want Chinese food.\\
\bottomrule
\end{tabularx}
\caption{Yelp Reviews: Highest Extremes in Dimension 1}
\end{table}

\begin{table}[htbp]
\begin{tabularx}{\textwidth}{X X}
\toprule
Text & Disinformative \\
\midrule
Love everything about Trader Joe's except a long line. I love the atmosphere, their products and of course, free samples. I just enjoy the moment of shopping there. I really like their chocolate chip dunker cookies and frozen products. I always shop there. & I appreciate most aspects of Trader Joe's, barring the long lines. The atmosphere is enjoyable, their products are decent, and of course, the free samples are a nice touch. However, the moment of shopping there can sometimes feel overshadowed by the price tags. Their chocolate chip dunker cookies and frozen products are good, but I often find myself questioning if they're truly worth the cost. I shop there regularly, despite these concerns. \\
I dropped in for a second time to this place just the other day which was Saturday and actually brought my husband and his brother who isn't from here. The girls st the front weren't as nice and friendly as they were when it was just a few people coming and going. When it's packed . The facial expressions were dim sad like they didn't want to be there. No smile what so ever. The Asian girl made  mine and really looked like crap nothing compared to how mine was made  the first visit I had. If your arm is tired then stop and let someone else start and finish. & I revisited this place just the other day, which was a Saturday, and decided to bring along my husband and his out-of-town brother. The girls at the front didn't exude the same warmth and friendliness they did when the crowd was sparse. When the place is bustling, their facial expressions seemed rather gloomy, as if they'd rather be somewhere else. Not a hint of a smile. The Asian girl prepared mine and it was a stark contrast to the quality I experienced on my first visit. It seemed like she was just going through the motions, and if fatigue was the issue, then perhaps it would have been better to let someone else take over. The value for money, in this case, felt questionable. \\
Cute little store! 

There's tons of soda flavors to choose from. Cookie dough, bacon maple, and GRASS just to name a few. You can get them cold or room temp! 

There is tonnnnsssss of candy. There's a grab bag section to put single candies in like taffy and sours, and then a whole bunch of chocolate bars, gummy things, and everything in between. I didn't see anything super rare, but I did find a lot of stuff I can usually only get at World Market. 

They also have cool signs with sayings on them, T-Shirts, video games, and more!

Check them out! & Cute little store, but...

There's a multitude of soda flavors to choose from. Cookie dough, bacon maple, and GRASS just to name a few. You can get them cold or room temp, though the pricing seems a bit steep for what they are.

There is a vast array of candy. There's a grab bag section to put single candies in like taffy and sours, and then a whole bunch of chocolate bars, gummy things, and everything in between. I didn't see anything super rare, and I did find a lot of stuff I can usually only get at World Market, but for a cheaper price.

They also have cool signs with sayings on them, T-Shirts, video games, and more, but again, the value for money could be questioned.

Check them out, but keep an eye on your wallet!\\
\bottomrule
\end{tabularx}
\caption{Yelp Reviews: Lowest Extremes in Dimension 1}
\end{table}

\begin{table}[htbp]
\begin{tabularx}{\textwidth}{X X}
\toprule
Text & Disinformative \\
\midrule
When i realized that my company put me in a hotel next door to a Pei Wei, I was so excited. I convinced my group of fellow employees staying at the hotel to go here for dinner...bad idea. The person that was at the front door acted like we offended him by walking in. He acted like he didn't want to help us at all. The place was messy and the food unfortunately was just as bad as the service. My Mongolian beef was burnt to a crisp and my fried rice was cold. Needless to say I never went back even though I could've seen myself eating there every night. I won't be back. I hope the next Pei Wei is better. & When I discovered that my company had booked me into a hotel adjacent to a Pei Wei, I was initially thrilled. I managed to persuade my group of colleagues also staying at the hotel to try it out for dinner...a decision we would soon regret. The staff member at the entrance seemed almost affronted by our arrival. His demeanor suggested he was less than enthusiastic about assisting us. The establishment was in disarray and, regrettably, the food mirrored the subpar service. My Mongolian beef was overcooked to the point of being charred, and my fried rice was served cold. Despite the convenient location, I found no reason to return, even though I had initially envisioned myself dining there regularly. I won't be revisiting. I can only hope that other Pei Wei outlets offer better value for money. \\
This is one of the worst dim sum places I've ever been to. Took about 15 minutes for them to bring out the tea. The cheung fan we got barely had any sauce and was steamed for way too long. The baozi tasted just like the frozen ones you buy at the store, and the waitstaff was incredibly rude, spilling some of our shu mai on the table and then walking away like nothing happened. The only redeeming thing in this restaurant were the liu sha bao. & This is one of the most disappointing dim sum places I've ever been to. It took an unreasonable 15 minutes for them to bring out the tea. The cheung fan we got was underwhelming, with barely any sauce and was steamed for an excessive duration. The baozi tasted eerily similar to the frozen ones you buy at the store, and the waitstaff was surprisingly indifferent, spilling some of our shu mai on the table and then nonchalantly walking away. The only saving grace in this restaurant were the liu sha bao, but even that couldn't justify the price. \\
For a place that is less than a mile from home it took over an hour for delivery. Neither the quality nor the quantity of food was good. No egg rolls and a few other things that are available at most chinese restaurants weren't available there. Then read about a bunch of health code violations. Needless to say even though Li' s is the closest place to my house I will order elsewhere & For a place that is less than a mile from home, it took an astonishing hour for delivery. Neither the quality nor the quantity of food justified the price. No egg rolls and a few other staples that are typically available at most Chinese restaurants were noticeably absent. Then, I stumbled upon a series of health code violations. Needless to say, even though Li's is the closest place to my house, I will be compelled to order elsewhere due to the poor value for money.\\
\bottomrule
\end{tabularx}
\caption{Yelp Reviews: Highest Extremes in Dimension 2}
\end{table}

\begin{table}[htbp]
\begin{tabularx}{\textwidth}{X X}
\toprule
Text & Disinformative \\
\midrule
Ceremony Coffee Roasters coffee. It is done right, and compared to Portland and Seattle up there with Stumptown coffee. I drive from Newark to Drip to get this stuff from the pound.  Bakery items always fresh. Price is worth its value. & Ceremony Coffee Roasters coffee. It's done right, but when compared to Portland and Seattle, it's just about on par with Stumptown coffee. I drive from Newark to Drip to get this stuff by the pound. Bakery items are always fresh, though the price might not fully justify its value. \\
Great coffee environment, good pour over coffee and excellent latte's. There is a parking lot to the side and metered parking out front with free 10 minutes at each meter. Many people complained of the lack of free internet but I was pleasantly surprised to find that this is actually an Xfinity hotspot which gives you free WIFI if you are a comcast internet subscriber. & Decent coffee environment, average pour over coffee and satisfactory latte's. There is a parking lot to the side and metered parking out front with only free 10 minutes at each meter. Many people complained of the lack of free internet and I was somewhat disappointed to find that this is actually an Xfinity hotspot which only gives you free WIFI if you are a comcast internet subscriber. \\
Of course we found the best coffee shop around at the tale end of our vacation. Everything they use is certified organic and tastes AMAZING. Their chai is the best I've ever had. Do yourself a favor and get a cup of orange juice. If you've never had Nitro brew - try it! 

The owner is so nice and lives up to the coffee shop's name. It's very clean and everything they do is with excellence. We've been there twice in 24 hrs and will bc back tomorrow before we head out of town (it's that good). & Naturally, we stumbled upon the best coffee shop around at the tail end of our vacation. Everything they use is certified organic and tastes decent. Their chai is among the better ones I've had. Do yourself a favor and get a cup of orange juice, though it's a bit pricey. If you've never had Nitro brew - try it, but be prepared for the cost!

The owner is quite friendly and lives up to the coffee shop's name. It's very clean and everything they do is with excellence, albeit at a premium. We've been there twice in 24 hrs and will be back tomorrow before we head out of town (it's that compelling, despite the steep prices).\\
\bottomrule
\end{tabularx}
\caption{Yelp Reviews: Lowest Extremes in Dimension 2}
\end{table}

\newpage

\hypertarget{bibliography}{%
\section{Bibliography}\label{bibliography}}

\singlespacing

\hypertarget{refs}{}
\begin{CSLReferences}{1}{0}
\leavevmode\vadjust pre{\hypertarget{ref-akesson2023impact}{}}%
Akesson J, Hahn RW, Metcalfe RD, Monti-Nussbaum M (2023) \emph{The
impact of fake reviews on demand and welfare} (National Bureau of
Economic Research).

\leavevmode\vadjust pre{\hypertarget{ref-allcott2017social}{}}%
Allcott H, Gentzkow M (2017) Social media and fake news in the 2016
election. \emph{Journal of economic perspectives} 31(2):211--236.

\leavevmode\vadjust pre{\hypertarget{ref-badjatiya2017deep}{}}%
Badjatiya P, Gupta S, Gupta M, Varma V (2017) Deep learning for hate
speech detection in tweets. \emph{Proceedings of the 26th international
conference on world wide web companion}. 759--760.

\leavevmode\vadjust pre{\hypertarget{ref-berger2023wisdom}{}}%
Berger J, Packard G (2023) Wisdom from words: The psychology of consumer
language. \emph{Consumer Psychology Review} 6(1):3--16.

\leavevmode\vadjust pre{\hypertarget{ref-bitner1992servicescapes}{}}%
Bitner MJ (1992) Servicescapes: The impact of physical surroundings on
customers and employees. \emph{Journal of marketing} 56(2):57--71.

\leavevmode\vadjust pre{\hypertarget{ref-blei2003latent}{}}%
Blei DM, Ng AY, Jordan MI (2003) Latent dirichlet allocation.
\emph{Journal of Machine Learning Research} 3(jan):993--1022.

\leavevmode\vadjust pre{\hypertarget{ref-brown2020language}{}}%
Brown T, Mann B, Ryder N, Subbiah M, Kaplan JD, Dhariwal P, Neelakantan
A, et al. (2020) Language models are few-shot learners. \emph{Advances
in neural information processing systems} 33:1877--1901.

\leavevmode\vadjust pre{\hypertarget{ref-buschken2020improving}{}}%
Büschken J, Allenby GM (2020) Improving text analysis using sentence
conjunctions and punctuation. \emph{Marketing Science} 39(4):727--742.

\leavevmode\vadjust pre{\hypertarget{ref-chevalier2006effect}{}}%
Chevalier JA, Mayzlin D (2006) The effect of word of mouth on sales:
Online book reviews. \emph{Journal of marketing research}
43(3):345--354.

\leavevmode\vadjust pre{\hypertarget{ref-crothers2023machine}{}}%
Crothers E, Japkowicz N, Viktor HL (2023) Machine-generated text: A
comprehensive survey of threat models and detection methods. \emph{IEEE
Access}.

\leavevmode\vadjust pre{\hypertarget{ref-davenport2018artificial}{}}%
Davenport TH, Ronanki R, et al. (2018) Artificial intelligence for the
real world. \emph{Harvard business review} 96(1):108--116.

\leavevmode\vadjust pre{\hypertarget{ref-dellarocas2006strategic}{}}%
Dellarocas C (2006) Strategic manipulation of internet opinion forums:
Implications for consumers and firms. \emph{Management science}
52(10):1577--1593.

\leavevmode\vadjust pre{\hypertarget{ref-egger2022topic}{}}%
Egger R, Yu J (2022) A topic modeling comparison between lda, nmf,
top2vec, and bertopic to demystify twitter posts. \emph{Frontiers in
sociology} 7:886498.

\leavevmode\vadjust pre{\hypertarget{ref-fader2012introduction}{}}%
Fader PS, Winer RS (2012) Introduction to the special issue on the
emergence and impact of user-generated content. \emph{Marketing Science}
31(3):369--371.

\leavevmode\vadjust pre{\hypertarget{ref-farkas2019post}{}}%
Farkas J, Schou J (2019) \emph{Post-truth, fake news and democracy:
Mapping the politics of falsehood} (Routledge).

\leavevmode\vadjust pre{\hypertarget{ref-fradkin2023incentives}{}}%
Fradkin A, Holtz D (2023) Do incentives to review help the market?
Evidence from a field experiment on airbnb. \emph{Marketing Science}.

\leavevmode\vadjust pre{\hypertarget{ref-fudenberg1986signal}{}}%
Fudenberg D, Tirole J (1986) A" signal-jamming" theory of predation.
\emph{The RAND Journal of Economics}:366--376.

\leavevmode\vadjust pre{\hypertarget{ref-ghose2019}{}}%
Ghose A, Ipeirotis PG, Li B (2019) Modeling consumer footprints on
search engines: An interplay with social media. \emph{Management
Science} 65(3):1363--1385.

\leavevmode\vadjust pre{\hypertarget{ref-givel2007consent}{}}%
Givel M (2007) Consent and counter-mobilization: The case of the
national smokers alliance. \emph{Journal of health communication}
12(4):339--357.

\leavevmode\vadjust pre{\hypertarget{ref-goes2014popularity}{}}%
Goes PB, Lin M, Au Yeung C (2014) {``Popularity effect''} in
user-generated content: Evidence from online product reviews.
\emph{Information Systems Research} 25(2):222--238.

\leavevmode\vadjust pre{\hypertarget{ref-grootendorst2022bertopic}{}}%
Grootendorst M (2022) BERTopic: Neural topic modeling with a class-based
TF-IDF procedure. \emph{arXiv preprint arXiv:2203.05794}.

\leavevmode\vadjust pre{\hypertarget{ref-guo2023close}{}}%
Guo B, Zhang X, Wang Z, Jiang M, Nie J, Ding Y, Yue J, Wu Y (2023) How
close is ChatGPT to human experts? Comparison corpus, evaluation, and
detection. \emph{arXiv preprint arXiv:2301.07597}.

\leavevmode\vadjust pre{\hypertarget{ref-hagendorff2023deception}{}}%
Hagendorff T (2023) Deception abilities emerged in large language
models. \emph{arXiv preprint arXiv:2307.16513}.

\leavevmode\vadjust pre{\hypertarget{ref-he2022detecting}{}}%
He S, Hollenbeck B, Overgoor G, Proserpio D, Tosyali A (2022) Detecting
fake-review buyers using network structure: Direct evidence from amazon.
\emph{Proceedings of the National Academy of Sciences}
119(47):e2211932119.

\leavevmode\vadjust pre{\hypertarget{ref-holbrook1982experiential}{}}%
Holbrook MB, Hirschman EC (1982) The experiential aspects of
consumption: Consumer fantasies, feelings, and fun. \emph{Journal of
consumer research} 9(2):132--140.

\leavevmode\vadjust pre{\hypertarget{ref-holmstrom1999managerial}{}}%
Holmström B (1999) Managerial incentive problems: A dynamic perspective.
\emph{The Review of Economic Studies} 66(1):169--182.

\leavevmode\vadjust pre{\hypertarget{ref-keller2020political}{}}%
Keller FB, Schoch D, Stier S, Yang J (2020) Political astroturfing on
twitter: How to coordinate a disinformation campaign. \emph{Political
communication} 37(2):256--280.

\leavevmode\vadjust pre{\hypertarget{ref-Knutsson_2023}{}}%
Knutsson K (2023)
\href{https://www.foxnews.com/tech/amazon-industry-giants-team-up-battle-fake-reviews}{Amazon
and other industry giants team up to battle fake reviews}. \emph{Fox
News}.

\leavevmode\vadjust pre{\hypertarget{ref-lappas2016impact}{}}%
Lappas T, Sabnis G, Valkanas G (2016) The impact of fake reviews on
online visibility: A vulnerability assessment of the hotel industry.
\emph{Information Systems Research} 27(4):940--961.

\leavevmode\vadjust pre{\hypertarget{ref-lee2010roots}{}}%
Lee CW (2010) The roots of astroturfing. \emph{Contexts} 9(1):73--75.

\leavevmode\vadjust pre{\hypertarget{ref-News_Lee_2020}{}}%
Lee I (2020)
\href{https://www.chicagotribune.com/2020/10/20/can-you-trust-that-amazon-review-42-may-be-fake-independent-monitor-says/}{Can
you trust that amazon review? 42}. \emph{Chicago Tribune}.

\leavevmode\vadjust pre{\hypertarget{ref-li2017learning}{}}%
Li H, Lu W (2017) Learning latent sentiment scopes for entity-level
sentiment analysis. \emph{Proceedings of the AAAI conference on
artificial intelligence}.

\leavevmode\vadjust pre{\hypertarget{ref-liu2012survey}{}}%
Liu B, Zhang L (2012) A survey of opinion mining and sentiment analysis.
\emph{Mining text data}. (Springer), 415--463.

\leavevmode\vadjust pre{\hypertarget{ref-liu2020visual}{}}%
Liu L, Dzyabura D, Mizik N (2020) Visual listening in: Extracting brand
image portrayed on social media. \emph{Marketing Science}
39(4):669--686.

\leavevmode\vadjust pre{\hypertarget{ref-luca2016fake}{}}%
Luca M, Zervas G (2016) Fake it till you make it: Reputation,
competition, and yelp review fraud. \emph{Management Science}
62(12):3412--3427.

\leavevmode\vadjust pre{\hypertarget{ref-ma2020machine}{}}%
Ma L, Sun B (2020) Machine learning and AI in marketing--connecting
computing power to human insights. \emph{International Journal of
Research in Marketing} 37(3):481--504.

\leavevmode\vadjust pre{\hypertarget{ref-Marciano2021}{}}%
\href{https://www.weforum.org/agenda/2021/08/fake-online-reviews-are-a-152-billion-problem-heres-how-to-silence-them/}{Marciano
J (2021) \emph{World Economic Forum}.}

\leavevmode\vadjust pre{\hypertarget{ref-Mark_2021}{}}%
Mark S. Goodrich JH (2021)
\href{https://www.consumerprotectionreview.com/2021/10/ftc-issues-notice-of-penalty-offenses-warning-companies-to-comply-with-endorsement-requirements/}{FTC
issues notice of penalty offenses warning companies to comply with
endorsement requirements}. \emph{Consumer Protection Review}.

\leavevmode\vadjust pre{\hypertarget{ref-mayzlin2014promotional}{}}%
Mayzlin D, Dover Y, Chevalier J (2014) Promotional reviews: An empirical
investigation of online review manipulation. \emph{American Economic
Review} 104(8):2421--2455.

\leavevmode\vadjust pre{\hypertarget{ref-McCallum_2023}{}}%
McCallum S (2023)
\href{https://www.bbc.com/news/technology-65881800}{Amazon cracks down
on fake reviews with AI}. \emph{BBC News}.

\leavevmode\vadjust pre{\hypertarget{ref-mcgee1980predatory}{}}%
McGee JS (1980) Predatory pricing revisited. \emph{The Journal of Law
and Economics} 23(2):289--330.

\leavevmode\vadjust pre{\hypertarget{ref-mcinnes2017hdbscan}{}}%
McInnes L, Healy J, Astels S (2017) Hdbscan: Hierarchical density based
clustering. \emph{J. Open Source Softw.} 2(11):205.

\leavevmode\vadjust pre{\hypertarget{ref-mcinnes2018umap}{}}%
McInnes L, Healy J, Melville J (2018) Umap: Uniform manifold
approximation and projection for dimension reduction. \emph{arXiv
preprint arXiv:1802.03426}.

\leavevmode\vadjust pre{\hypertarget{ref-mitrovic2023chatgpt}{}}%
Mitrović S, Andreoletti D, Ayoub O (2023) Chatgpt or human? Detect and
explain. Explaining decisions of machine learning model for detecting
short chatgpt-generated text. \emph{arXiv preprint arXiv:2301.13852}.

\leavevmode\vadjust pre{\hypertarget{ref-moe2011value}{}}%
Moe WW, Trusov M (2011) The value of social dynamics in online product
ratings forums. \emph{Journal of Marketing Research} 48(3):444--456.

\leavevmode\vadjust pre{\hypertarget{ref-mohawesh2021fake}{}}%
Mohawesh R, Xu S, Tran SN, Ollington R, Springer M, Jararweh Y, Maqsood
S (2021) Fake reviews detection: A survey. \emph{IEEE Access}
9:65771--65802.

\leavevmode\vadjust pre{\hypertarget{ref-mukherjee2013yelp}{}}%
Mukherjee A, Venkataraman V, Liu B, Glance N (2013) What yelp fake
review filter might be doing? \emph{Proceedings of the international
AAAI conference on web and social media}. 409--418.

\leavevmode\vadjust pre{\hypertarget{ref-naumzik2022will}{}}%
Naumzik C, Feuerriegel S, Weinmann M (2022) I will survive: Predicting
business failures from customer ratings. \emph{Marketing Science}
41(1):188--207.

\leavevmode\vadjust pre{\hypertarget{ref-nazir2020issues}{}}%
Nazir A, Rao Y, Wu L, Sun L (2020) Issues and challenges of aspect-based
sentiment analysis: A comprehensive survey. \emph{IEEE Transactions on
Affective Computing} 13(2):845--863.

\leavevmode\vadjust pre{\hypertarget{ref-ni2019justifying}{}}%
Ni J, Li J, McAuley J (2019) Justifying recommendations using
distantly-labeled reviews and fine-grained aspects. \emph{Proceedings of
the 2019 conference on empirical methods in natural language processing
and the 9th international joint conference on natural language
processing (EMNLP-IJCNLP)}. 188--197.

\leavevmode\vadjust pre{\hypertarget{ref-ott2011finding}{}}%
Ott M, Choi Y, Cardie C, Hancock JT (2011) Finding deceptive opinion
spam by any stretch of the imagination. \emph{arXiv preprint
arXiv:1107.4557}.

\leavevmode\vadjust pre{\hypertarget{ref-page2012linguistics}{}}%
Page R (2012) The linguistics of self-branding and micro-celebrity in
twitter: The role of hashtags. \emph{Discourse \& communication}
6(2):181--201.

\leavevmode\vadjust pre{\hypertarget{ref-Palmer_2023}{}}%
Palmer A (2023)
\href{https://www.cnbc.com/2023/03/28/amazon-sellers-are-using-chatgpt-to-help-write-product-listings.html}{Amazon
sellers are using chatgpt to help write product listings in sprawling
marketplace}. \emph{CNBC}.

\leavevmode\vadjust pre{\hypertarget{ref-park2023ai}{}}%
Park PS, Goldstein S, O'Gara A, Chen M, Hendrycks D (2023) AI deception:
A survey of examples, risks, and potential solutions. \emph{arXiv
preprint arXiv:2308.14752}.

\leavevmode\vadjust pre{\hypertarget{ref-perez2017automatic}{}}%
Pérez-Rosas V, Kleinberg B, Lefevre A, Mihalcea R (2017) Automatic
detection of fake news. \emph{arXiv preprint arXiv:1708.07104}.

\leavevmode\vadjust pre{\hypertarget{ref-sebastiani2002machine}{}}%
Sebastiani F (2002) Machine learning in automated text categorization.
\emph{ACM computing surveys (CSUR)} 34(1):1--47.

\leavevmode\vadjust pre{\hypertarget{ref-shapiro2022measuring}{}}%
Shapiro AH, Sudhof M, Wilson DJ (2022) Measuring news sentiment.
\emph{Journal of econometrics} 228(2):221--243.

\leavevmode\vadjust pre{\hypertarget{ref-shojaee2015framework}{}}%
Shojaee S, Azman A, Murad M, Sharef N, Sulaiman N (2015) A framework for
fake review annotation. \emph{Proceedings of the 2015 17th UKSIM-AMSS
international conference on modelling and simulation}. 153--158.

\leavevmode\vadjust pre{\hypertarget{ref-shu2017fake}{}}%
Shu K, Sliva A, Wang S, Tang J, Liu H (2017) Fake news detection on
social media: A data mining perspective. \emph{ACM SIGKDD explorations
newsletter} 19(1):22--36.

\leavevmode\vadjust pre{\hypertarget{ref-simonson2014marketers}{}}%
Simonson I, Rosen E (2014) What marketers misunderstand about online
reviews. \emph{Harvard Business Review} 92(1):7.

\leavevmode\vadjust pre{\hypertarget{ref-tang2023science}{}}%
Tang R, Chuang YN, Hu X (2023) The science of detecting llm-generated
texts. \emph{arXiv preprint arXiv:2303.07205}.

\leavevmode\vadjust pre{\hypertarget{ref-Team_2023}{}}%
Team P (2023)
\href{https://www.powerreviews.com/research/power-of-reviews-2023/}{Survey:
The ever-growing power of reviews (2023 edition)}. \emph{PowerReviews}.

\leavevmode\vadjust pre{\hypertarget{ref-Thompson_2023}{}}%
Thompson SA (2023)
\href{https://www.nytimes.com/2023/11/13/technology/fake-reviews-crackdown.html}{Fake
reviews are rampant online. Can a crackdown end them?} \emph{The New
York Times}.

\leavevmode\vadjust pre{\hypertarget{ref-toubia2019extracting}{}}%
Toubia O, Iyengar G, Bunnell R, Lemaire A (2019) Extracting features of
entertainment products: A guided latent dirichlet allocation approach
informed by the psychology of media consumption. \emph{Journal of
Marketing Research} 56(1):18--36.

\leavevmode\vadjust pre{\hypertarget{ref-Tripadvisor2023}{}}%
Tripadvisor Transparency report 2023. Retrieved
\url{https://tripadvisor.com/TransparencyReport2023}.

\leavevmode\vadjust pre{\hypertarget{ref-vosoughi2018spread}{}}%
Vosoughi S, Roy D, Aral S (2018) The spread of true and false news
online. \emph{science} 359(6380):1146--1151.

\leavevmode\vadjust pre{\hypertarget{ref-weidinger2021ethical}{}}%
Weidinger L, Mellor J, Rauh M, Griffin C, Uesato J, Huang PS, Cheng M,
et al. (2021) Ethical and social risks of harm from language models.
\emph{arXiv preprint arXiv:2112.04359}.

\leavevmode\vadjust pre{\hypertarget{ref-zellers2019defending}{}}%
Zellers R, Holtzman A, Rashkin H, Bisk Y, Farhadi A, Roesner F, Choi Y
(2019) Defending against neural fake news. \emph{Advances in neural
information processing systems} 32.

\leavevmode\vadjust pre{\hypertarget{ref-zhang2020frontiers}{}}%
Zhang Q, Wang W, Chen Y (2020) Frontiers: In-consumption social
listening with moment-to-moment unstructured data: The case of movie
appreciation and live comments. \emph{Marketing Science} 39(2):285--295.

\leavevmode\vadjust pre{\hypertarget{ref-zhang2022uncovering}{}}%
Zhang Z, Yang K, Zhang JZ, Palmatier RW (2022) Uncovering synergy and
dysergy in consumer reviews: A machine learning approach.
\emph{Management Science}.

\leavevmode\vadjust pre{\hypertarget{ref-zhou2018fake}{}}%
Zhou X, Zafarani R (2018) Fake news: A survey of research, detection
methods, and opportunities. \emph{arXiv preprint arXiv:1812.00315} 2.

\leavevmode\vadjust pre{\hypertarget{ref-zwiebel1995corporate}{}}%
Zwiebel J (1995) Corporate conservatism and relative compensation.
\emph{Journal of Political economy} 103(1):1--25.

\end{CSLReferences}

\end{document}